\DocumentMetadata{}
\documentclass[sigconf]{acmart} 
\usepackage{makecell}
\usepackage{algorithm}
\usepackage{algpseudocode}
\usepackage{amsmath}
\usepackage{algorithm,algpseudocode}
\usepackage{hyperref}
\usepackage[normalem]{ulem}
\usepackage{xspace}

\makeatletter
\def\@ACM@checkaffil{
    \if@ACM@instpresent\else
    \ClassWarningNoLine{\@classname}{No institution present for an affiliation}%
    \fi
    \if@ACM@citypresent\else
    \ClassWarningNoLine{\@classname}{No city present for an affiliation}%
    \fi
    \if@ACM@countrypresent\else
        \ClassWarningNoLine{\@classname}{No country present for an affiliation}%
    \fi
}
\makeatother

\newcommand{\rev}[1]{\textcolor{black}{#1}}

\definecolor{bcolor1}{rgb}     {1.0,0.0,0.0}
\definecolor{darkgreen}{rgb}     {0.0,0.5,0.0}
\definecolor{blue}{rgb}     {0,0.0,1.0}

\definecolor{red}{rgb}{1, 0, 0}
\definecolor{black}{rgb}{0, 0, 0}
\definecolor{blue}{rgb}{0, 0, 1}

\algnewcommand{\Inputs}[1]{%
  \State \textbf{Inputs:}
  \Statex \hspace*{\algorithmicindent}\parbox[t]{.8\linewidth}{\raggedright #1}
}
\algnewcommand{\Outputs}[1]{%
  \State \textbf{Outputs:}
  \Statex \hspace*{\algorithmicindent}\parbox[t]{.8\linewidth}{\raggedright #1}
}
\algnewcommand{\Initialize}[1]{%
  \State \textbf{Initialize:}
  \Statex \hspace*{\algorithmicindent}\parbox[t]{.8\linewidth}{\raggedright #1}
}

\AtBeginDocument{%
  \providecommand\BibTeX{{%
    \normalfont B\kern-0.5em{\scshape i\kern-0.25em b}\kern-0.8em\TeX}}}
    
\newcommand{\concept}{HOMI\xspace}
\newcommand{\conceptnospace}{HOMI}
\newcommand{\method}{NAF$^+$\xspace}
\newcommand{\methodnospace}{NAF$^+$}

\makeatletter
\def\@ACM@copyright@check@cc{}
\makeatother
\copyrightyear{2026}
\acmYear{2026}
\setcopyright{cc}
\setcctype{by}
\acmConference[CHI '26]{Proceedings of the 2026 CHI Conference on Human Factors in Computing Systems}{April 13--17, 2026}{Barcelona, Spain}
\acmBooktitle{Proceedings of the 2026 CHI Conference on Human Factors in Computing Systems (CHI '26), April 13--17, 2026, Barcelona, Spain}
\acmPrice{}
\acmDOI{10.1145/3772318.3791976}
\acmISBN{979-8-4007-2278-3/2026/04}

\begin{document}


\title[Efficient Human-in-the-Loop Optimization via Priors Learned from User Models]{Efficient Human-in-the-Loop Optimization via \\ Priors Learned from User Models}






\author{Yi-Chi Liao}
\orcid{0000-0002-2670-8328}
\affiliation{%
  \institution{Saarland University, Saarland Informatics Campus, Saarbrücken}
  \country{Germany}
}
\affiliation{%
  \institution{ETH Zurich, Zurich}
  \country{Switzerland}
}
\email{yichi.liao@inf.ethz.ch}

\author{João Belo}
\orcid{0000-0002-3403-2970}
\affiliation{%
  \institution{Saarland University, Saarland Informatics Campus, Saarbrücken}
  \country{Germany}
}
\email{jbelo@cs.uni-saarland.de}

\author{Hee-Seung Moon}
\orcid{0000-0003-0882-2335}
\affiliation{%
  \institution{Chung-Ang University, Seoul}
  \country{Republic of Korea}
}
\email{hsmoon@cau.ac.kr}

\author{Jürgen Steimle}
\orcid{0000-0003-3493-8745}
\affiliation{%
  \institution{Saarland University, Saarland Informatics Campus, Saarbrücken}
  \country{Germany}
}
\email{steimle@cs.uni-saarland.de}

\author{Anna Maria Feit}
\orcid{0000-0003-4168-6099}
\affiliation{%
  \institution{Saarland University, Saarland Informatics Campus, Saarbrücken}
  \country{Germany}
}
\email{feit@cs.uni-saarland.de}

\renewcommand{\shortauthors}{Liao et al.}


\begin{abstract}
Human-in-the-loop optimization identifies optimal interface designs by iteratively observing user performance. 
However, it often requires numerous iterations due to the lack of prior information. 
While recent approaches have accelerated this process by leveraging previous optimization data, collecting user data remains costly and often impractical. 
We present a conceptual framework, \emph{Human-in-the-Loop Optimization with Model-Informed Priors} (HOMI), which augments human-in-the-loop optimization with a training phase where the optimizer learns adaptation strategies from diverse, synthetic user data generated with predictive models before deployment. 
To realize HOMI, we introduce Neural Acquisition Function$^+$ (NAF$^+$), a Bayesian optimization method featuring a neural acquisition function trained with reinforcement learning. 
NAF$^+$ learns optimization strategies from large-scale synthetic data, improving efficiency in real-time optimization with users. 
We evaluate HOMI and NAF$^+$ with mid-air keyboard optimization, a representative VR input task. 
Our work presents a new approach for more efficient interface adaptation by bridging \emph{in situ} and \emph{in silico} optimization processes.

\end{abstract}






\maketitle

\begin{figure*}[t]
\centering
\includegraphics[width=1\textwidth]{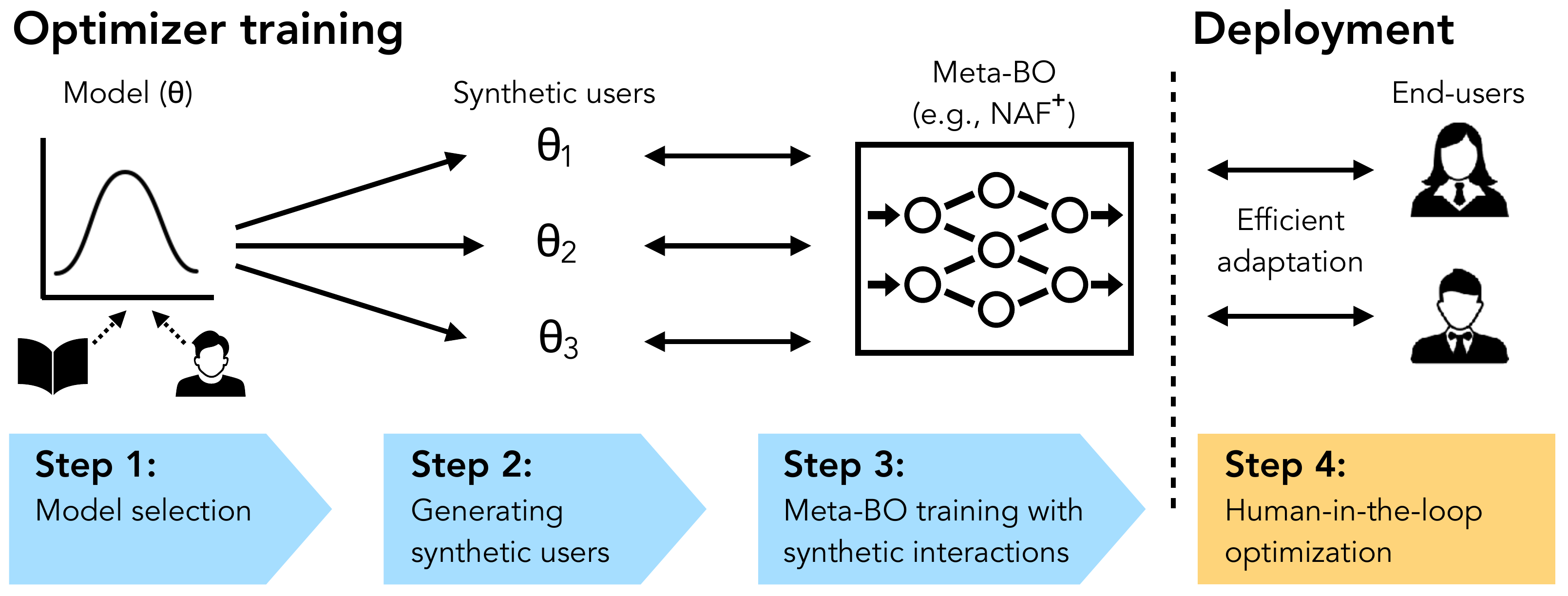}
  \caption{
  The general steps of our \concept framework. \textbf{Step 1: Model selection} — Designers or developers select a model, parameterized by $\theta$, that is relevant to the target task, and optionally fit the model's parameters (e.g., Fitts' law, with parameters $a$ and $b$) using minimal user data to better reflect the target context. \textbf{Step 2: Synthetic user generation} — A diverse set of synthetic users is created by sampling different parameter settings ${\theta_1, \theta_2, \theta_3, \dots}$ from the fitted model's parameter distribution. \textbf{Step 3: Meta-BO training} — The meta-optimizer interacts extensively with these synthetic users to learn efficient strategies for optimizing across user variability. \textbf{Step 4: Deployment} — The trained meta-BO is deployed with real users and quickly adapts to individual performance or preferences using the learned prior experience.
  }
 ~\label{fig:concept}
\end{figure*}

\section{Introduction}

Efficient and effective interface optimization for different users and contexts is a long-standing goal in human-computer interaction (HCI)~\cite{norcio1989adaptive, zhai2002performance, rao2019engineering, lavie2010benefits}. 
Model-based optimization approaches address this goal by employing computational optimizers to search for interface designs that maximize predicted user performance, based on established behavioral models.
However, most such models capture only population-level behavior, limiting their ability to support personalized interface optimization.
Although recent work has attempted to optimize interfaces for individuals or contexts by inferring user-specific or context-specific model parameters during use, this inference is often challenging or time-consuming~\cite{beaumont2002approximate, moon2023amortized}, making it difficult to scale or deploy in real-time settings.
In recent years, human-in-the-loop optimization (HILO) emerged as a more practical alternative \cite{10.1145/3313831.3376244, 10.1145/2858036.2858253, 10.1145/3386569.3392409, 10.1371/journal.pone.0184054}. 
Unlike model-based approaches that rely on user models, HILO operates directly on actual users: the optimizer proposes a design candidate, users interact with it, their performance or feedback is evaluated, and the optimizer iteratively suggests new designs to improve the user experience.
A key advantage of this paradigm is that optimization is driven by users' real-time evaluations, enabling personalization without requiring a user-model inference step.
Various computational techniques, such as Bayesian optimization \cite{koyama2017sequential, 10024515, 10.5555/1921427.1921443, 10.1145/3290605.3300482}, evolutionary and genetic algorithms \cite{battiti2010brain, meng2008research}, and heuristic-based techniques \cite{10.1145/3313831.3376244} have been explored to realize HILO. 
Among these, Bayesian optimization (BO) stands out due to its better performance and minimal assumptions about the task \cite{frazier2018tutorial}.
HILO with BO has been applied in domains such as novel interaction techniques \cite{10.1145/3313831.3376262}, design tools \cite{10.1145/2858036.2858253, 10.5555/1921427.1921443}, and wearable interfaces \cite{10.1145/3613904.3642071, 10.1371/journal.pone.0184054}.

While effective for personalizing designs, HILO has a key limitation: the optimization process is often lengthy, typically requiring many rounds of user evaluations to identify a promising interface design~\cite{wang2022recent, pu2024adamo}.
The underlying challenge is that the optimizer usually begins with little or no task-specific information, requiring it to explore the design space with random design candidates before gradually converging toward an optimal solution.
Such prolonged adaptation not only exposes users to suboptimal designs, but also forces them to repeatedly adjust their behavior, or even re-learn interaction strategies as the interface continues to change.
This is especially problematic for interactions that demand instant and stable performance, such as most input techniques.
Ideally, an optimizer should therefore identify a promising, or even optimal, design within as few trials as possible, and then fix the interface to provide a stable experience for the remaining use.
Consequently, efficiently finding high-performing designs early has become a central goal in HILO and broad adaptive interfaces \cite{10.1145/3613904.3642071, koyama2017sequential}.

Recent work has attempted to enhance the adaptation efficiency by incorporating prior data from past users to initialize or guide the optimization process, a concept known as transfer learning or meta-learning.
For example, previous works have explored constructing a single surrogate model based on all the previous users' evaluation results \cite{doi:10.1177/26339137241241313}, storing a series of separate prior user models and aggregated them via a weighted-sum approach during deployment \cite{10.1145/3613904.3642071}, or maintaining an adaptive surrogate model using Bayesian neural networks to accumulate growing data \cite{liao2025continual}.
These works demonstrated that augmenting HILO with prior experience is a viable and general path toward solving the cold-start issue.
However, a key bottleneck remains: the reliance on real human data.
Prior transfer- or meta-learning–based methods require data from at least ten users to be effective \cite{doi:10.1177/26339137241241313, 10.1145/3613904.3642071}, which can be prohibitively expensive or impractical.
Moreover, any changes to the interaction task or interface require fresh rounds of data collection, severely limiting scalability.
This motivates a fundamental question:
\emph{Is it possible to train computational optimizers for higher online efficiency, yet, without relying on real human data?}

In this work, we propose a new computational framework called \textit{Human-in-the-Loop Optimization with Model-Informed Priors (\conceptnospace)}, illustrated in \autoref{fig:concept}. 
Here, ``model-informed priors'' refer to the learned optimization strategies acquired by training with predictive models.
Rather than relying on real user data for training the optimizer, HOMI leverages data from synthetic users simulated from these models.
This offline training phase allows Bayesian optimization to develop a broad understanding of the design space and to learn effective adaptation strategies across diverse user contexts before online optimization begins.
A key advantage of using model-informed priors is that the optimizer can be trained with a wide variety of synthetic users, enabling generalization across different usage scenarios.
In contrast, training with real users limits the optimizer's knowledge to a small number of specific individuals, reducing its adaptability.
At deployment, the trained optimizer switches to a human-in-the-loop mode, integrating prior knowledge with live user feedback to efficiently optimize the interface design in real time.

To realize the HOMI framework, we introduce a novel Bayesian optimization method enhanced by meta-learning; this type of approaches is commonly referred to as meta-BO in the machine learning field \cite{LI2021463, WANG202190}.
Our method, called Neural Acquisition Function$^+$ (NAF$^+$), builds on the work of \citet{volpp2019meta}, who proposed replacing acquisition functions in Bayesian optimization with neural networks trained via reinforcement learning \cite{kaelbling1996reinforcement, sutton2018reinforcement}.
Leveraging the scalability of neural networks, \method can be trained on large-scale synthetic user datasets, and incorporates two novel features tailored specifically for human-in-the-loop scenarios.
First, \method provides designers with enhanced flexibility in setting design objectives by dynamically aggregating multiple design objectives into a single weighted-sum objective for which individual weights can be flexibly tuned in practical applications.
Secondly, \method ensures robust performance across the diverse behaviors of end-users, even if their behavior varies significantly from that of synthetic users. Therefore, we propose a novelty detector that estimates how likely a user at run-time is different from the synthetic users, which is used to aggregate outputs from the neural and a standard acquisition function. 
We validate the effectiveness of our method through a series of synthetic tests, showing that \method converges more efficiently than established meta-BO approaches, adapts its optimization strategy to varying objective weightings, and remains robust when encountering novel synthetic users.

We further demonstrate the effectiveness of our \concept by applying \method to real-time adaptation of a mid-air keyboard for direct touch input, which is an essential VR text entry technique, in a study with real users.
The training involved synthetic users constructed via established models in HCI, including Fitts' law \cite{10.1207/s15327051hci0701_3} and typing error model \cite{10.1145/2984511.2984546}.
The results showed \concept enabled by \method outperforms the established baselines, including both manual keyboard setting and standard BO in that it significantly reduced the number of iterations required to  personalize the keyboard.

To summarize, we make the following contributions:
\begin{itemize}
    \item \textbf{Concept and Framework}: We propose \conceptnospace, a novel framework that trains HILO using synthetic user data generated from models, enabling scalable and efficient interface adaptation. Critically, \concept repositions the role of user models from being optimization targets to training resources.
    \item \textbf{Method}: We introduce \methodnospace, a Bayesian optimization method with a neural acquisition function trained on synthetic data, supporting dynamic objective weighting and robust generalization to unseen user behaviors.
    \item \textbf{Demonstration}: We evaluate HOMI and NAF$^+$ on mid-air keyboard adaptation, showing that our approach significantly outperforms standard BO and manual baselines in online adaptation scenarios.
\end{itemize}

\section{Background \& Related work}

We review the key topics that are relevant to this work, especially focusing on model-based optimization in HCI, HILO, and the techniques that augment Bayesian optimization with prior data for efficient interface adaptation.

\subsection{Model-based Optimization in HCI}

Optimization is an essential process in HCI and design, even if they are not explicitly articulated as such. 
In traditional user-centered design workflows, optimization is performed manually: designers generate a set of candidate designs, which are then evaluated through user testing and iteration \cite{abras2004user}.
The field of computational interaction~\cite{compintbook} aimed to integrate computational optimization methods in the design process to automatically identify optimal design candidates. 
This idea is not new: engineering fields have long used optimization workflows in which a solver proposes candidate designs, and these are evaluated through domain-specific models (e.g., physical, mechanical, chemical) to identify optimal solutions \cite{martins2021engineering}.
This approach has proven successful in fields such as mechanical engineering \cite{rao2011teaching}, chemical engineering \cite{bhaskar2000applications}, and architectural design \cite{michalek2002architectural}.
HCI researchers have adopted similar methods: deploying computational solvers over user models to identify optimal interaction designs.
Examples include menu optimization \cite{bailly2013menuoptimizer, todi2021adapting}, keyboard layout design \cite{feit2021azerty, karrenbauer2014improvements}, and user interface layout adaptation in AR/VR environments \cite{lindlbauer2019context, evangelista2022auit}, and more.

With the rapid growth of user modeling approaches, model-based optimization has shown success in generating promising design candidates.
A key opportunity can extend to enabling \emph{adaptive} interfaces: given a new user and context, in principle, one could first infer that user's model parameters (in other words, \emph{inverse modeling} or \emph{model inference}), then run a computational solver over that personalized model to derive an optimized interface.
However, this ideal pipeline is rarely feasible in practice.
Inverse modeling is computationally expensive and often ill-posed~\cite{kangasraasio2017inferring, moon2022speeding}: multiple parameter combinations can reproduce the same behavioral observations.
As a result, most model-based optimization in HCI has targeted an abstract \emph{average} user, or a fixed user group~\cite{sarcar2018ability, todi2021adapting}, and the deployed systems typically \emph{do not update or adapt their underlying model based on live observations}.

This paper circumvents this inverse-modeling bottleneck.
Instead of inferring exact model parameters for each user, we pursue \emph{human-in-the-loop adaptation} to directly identify the optimal interfaces for each individual with minimal user intervention.
The \concept framework builds on the foundation of model-based optimization, but reframes user models not as fixed targets to optimize against, but as generators of synthetic user data for training an optimizer.

\subsection{Human-in-the-Loop Bayesian Optimization}

Human-in-the-loop optimization (HILO) is an emerging optimization paradigm in HCI that does not rely on predefined user models.
Instead, HILO learns and optimizes directly from online observations, which may include user performance, preferences, or other forms of interaction feedback.
The objective function $f$ is usually a user-centered metric (e.g., completion time \cite{liao2025continual}, preferences \cite{10.1145/3386569.3392444}, etc.) that we want to optimize for \cite{10.1145/3313831.3376262, 10.1145/3613904.3642071, 10.1371/journal.pone.0184054, 10.1145/3526113.3545690}.
Each unique design candidate $x$ within the designated design space $\mathcal{X}$ results in a user-related objective function value, $f(x)$.
The goal of HILO is to identify the global optimal design $x^*$ that maximizes (or minimizes) the objective value $f(x^*)$.
Such optimization problems are challenging since the only way to access the user-relevant objective function is through costly user evaluations. 
Bayesian optimization (BO) is a popular computational method for human-in-the-loop optimization due to its higher sample-efficiency than other approaches (requiring fewer user evaluations), minimum requirements and assumptions of the problem, and reliable performances \cite{MOCKUS1975428, 7352306}. 
Within Bayesian optimization, a surrogate model, typically a Gaussian Process regression (GP), captures the properties of the target function.
This model is continually refined with online function observations (the result of user evaluation), enhancing its prediction accuracy.
An acquisition function utilizes the information of the surrogate model to compute the acquisition value (i.e., the worth or potential value) of a new candidate design. 
The candidate with the highest acquisition value is then selected for user evaluation in the subsequent iteration.

Bayesian optimization has found many human-involved applications where the sample efficiency is particularly important, such as exoskeleton systems~\cite{10.1371/journal.pone.0184054, kim2019bayesian, tucker2020human}, wearable interfaces~\cite{10024515, guo2019xgboost}, haptic displays~\cite{10024515,catkin2023preference}, input techniques~\cite{doi:10.1177/26339137241241313}, and design tools~\cite{10.5555/1921427.1921443, koyama2017sequential, 10.1145/3657643, Langerak2026CostAwareBO}.
Despite its robust performance and relatively better efficiency over other optimization approaches (simulated annealing, evolutionary algorithms, etc.), a key limitation of BO lies in its ``cold-start'' problem.
Since BO generally starts without any prior knowledge of the objective function, it must rely on random exploration in the early stages, which can slow down convergence.
To address this, recent work has explored incorporating transfer learning and meta-learning to accelerate the optimization process, which we will review in the next subsection.

\subsection{Meta-learning for BO}

Meta-learning \cite{WANG202190} is a machine learning paradigm that aims to accelerate learning on a novel task by leveraging experience from previous tasks with a similar structure.
This idea has also been extended to Bayesian Optimization (BO), where it is known as meta-BO: instead of optimizing each problem from scratch, the optimizer draws on past optimization tasks to guide search on a new one.
In our setting, a task corresponds to \emph{optimizing the interface for a specific user}, making meta-BO a natural fit, as it enables rapid adaptation when a new user arrives.
Recent surveys, such as \citet{bai2023transfer} provides a comprehensive overview of recent meta-BO methods.
A common approach is to merge all data from prior tasks into a single unified surrogate model, often implemented with multi-task Gaussian Processes (GPs), sparse GPs, or other implementations~\cite{10.5555/3042817.3042916, bonilla2007multi, NIPS2013_f33ba15e, yogatama2014efficient}.
This unified surrogate is then used to guide optimization for a new user.
Another line of work constructs separate surrogate models for each task, and combines their predictions through a weighted sum for a new task~\cite{li2022transbo, 10.1007/978-3-319-46128-1_3, wistuba2018scalable}.
These methods are conceptually simple, easy to implement, and have been applied in HCI contexts~\cite{10.1145/3613904.3642071, li2025efficient}.
However, both suffer from scalability issues as the number of past tasks grows: the unified model requires cubic-time GP inference, whereas the weighted-sum methods scale linearly.

In this work, we exploit the fact that user models can generate unlimited synthetic data, providing far richer information than prior data in typical meta-BO settings.
However, leveraging this large-scale dataset requires meta-BO methods that scale with the number of tasks and volume of data, which classical approaches cannot satisfy.
We therefore follow a third direction in meta-BO: replacing hand-crafted components of BO with deep neural networks~\cite{wistuba2021few, hsieh2021reinforced}, enabling efficient inference even with a large number of prior tasks.
Our \method is built upon \citet{volpp2019meta}, which replaces the manually crafted acquisition function with a deep neural network, called Neural Acquisition Function (NAF).
Pre-trained with data from optimization across users, NAF learns a general strategy to select the next design candidates promising to be evaluated. 
While individual users may differ in their behaviors or preferences to some extent, we assume an underlying shared structure across user-optimization tasks that meta-BO can learn and transfer.
Note that the acquisition function in \method is trained via reinforcement learning, which we describe in more detail in the next section.

\section{Human-in-the-Loop Optimization with Model-Informed Priors}

Our first contribution is the introduction of a novel conceptual framework, which we term \conceptnospace, and the corresponding steps of pretraining meta-BO with synthetic users.
While \method represents one possible realization, the \concept framework is intended to be general and extensible, enabling future work to explore alternative implementations.
Below, we outline the core components of the framework.

\subsection{Core Elements of \conceptnospace:}
\concept is an interactive framework built upon three key components: 
\textbf{meta-Bayesian optimization (meta-BO)}, \textbf{user models}, and \textbf{end-users}.
Within the \concept framework, \textbf{Meta-BO} refers to a computational optimizer trained across a distribution of synthetic users operating on the same interaction.
During the \textit{optimizer training phase}, meta-BO interacts with synthetic users generated from selected, relevant user models.
Since these interactions are simulated, the optimizer can perform an unlimited number of evaluations to learn generalized adaptation strategies.
\textbf{User models} are parameterized functions (e.g., Fitts' law) that simulate user behavior and performance.
By sampling different model parameter settings, we can generate a diverse population of synthetic users, each representing different motor abilities, interaction speeds, or usage contexts.
\textbf{End-users} are real human participants interacting with the system during the \textit{deployment phase}.
During this phase, meta-BO observes users' real-time feedback, such as performance metrics, interaction preferences, or other behavioral data, and uses it to rapidly infer user-specific characteristics and identify optimal designs with as few evaluation iterations as possible.

\subsection{General Steps in \conceptnospace:}

\autoref{fig:concept} illustrates the four main steps in the \concept framework.
Note that the first three steps comprise the offline meta-BO training phase, while only the fourth step occurs during real-time deployment.

\textbf{Step 1: Model selection.} 
\concept framework requires synthetic users with varying performances to train the optimizer. 
These synthetic users are generated by parameterized user models~\cite{purificato2024user}, where varying model parameters correspond to different user abilities, devices, or usage contexts.
For example, for typing and target selection, Fitt's law is the established model for predicting the movement time, which can be parametrized to predict performance with different devices or different abilities. 
The key novelty of \concept is leveraging these models to simulate a large range of typical user behaviors by systematically varying parameters across a range of values. 

For many interaction tasks, such parameter ranges are provided in the literature.
For pointing and target selection, prior work reports Fitts' law parameters across devices and conditions~\cite{soukoreff2004towards, mackenzie1992fitts}.
For touch input, studies provide Gaussian touch-error models with empirically validated parameters~\cite{10.1145/2501988.2502058, orphanides2017touchscreen}.
For text entry, typing models offer parameterized keystroke latencies and error distributions~\cite{soukoreff2003metrics, mackenzie2002model}.
Similarly, models for continuous motion, such as the Steering Law, include validated parameter sets~\cite{accot1997beyond}.
Together, these models allow researchers to construct diverse synthetic users by sampling from known parameter ranges or making informed parameter choices, without collecting new user data or refitting models from scratch.
In summary, the first step of \concept is to identify task-relevant models that capture how design variations affect user performance.

\textbf{Step 2: Synthetic user generation}: 
Given selected models and parameter ranges, synthetic users are generated by sampling model parameters from the corresponding distributions.
Each synthetic user represents a distinct set of user characteristics, leading to a diverse simulated population.

\textbf{Step 3: Meta-BO training with synthetic interactions}: Before deploying to the end-users, meta-BO interacts with these synthetic users to learn how to efficiently identify optimal designs across different user profiles.
Since synthetic evaluations are inexpensive compared to human evaluations, this phase allows for large-scale training.
The diversity of synthetic users further ensures the optimizer does not overfit to specific traits, but rather learns transferable strategies for personalization.

\textbf{Step 4: Human-in-the-loop optimization}: The trained optimizer, meta-BO, is deployed with real users, leveraging prior experience to adapt quickly and effectively to individual needs, ultimately addressing the cold-start problem.

\section{Neural Acquisition Function$^+$}

To realize the concept of \concept, we introduce a novel meta-BO method for HCI purposes: \textit{Neural Acquisition Function$^+$} (\methodnospace).
The central idea behind \method is to replace the manually designed acquisition function in Bayesian Optimization with a neural network, trained on synthetic user data. This enables the optimizer to learn efficient adaptation strategies prior to deployment, resulting in faster convergence during real-time human-in-the-loop optimization.
Our \method is built upon a key previous work, Neural Acquisition Function (NAF) \cite{volpp2019meta}, and extends it with two novel features that offer designers flexibility and ensure robust performance for real-world users.
In this section, we first introduce the necessary preliminaries: standard acquisition functions in BO, reinforcement learning (the core training method behind NAF), and the original NAF formulation.
We then present the core contributions and architectural enhancements of our proposed \methodnospace.

\subsection{Preliminaries}
\label{sec:pre}

\paragraph{1. Acquisition functions and surrogate models in BO}
BO has two core components: 
a surrogate model and an acquisition function.
The surrogate model fits the observed design points ($x$) and corresponding objective values ($y$), enabling prediction of the posterior mean and variance of the objective function at any candidate $x$.
The acquisition function utilizes the surrogate model's predictions to compute ``acquisition values'' across the design space, and the candidate with the highest value is selected for evaluation. 
EI is one of the most widely adopted acquisition functions in BO due to its simplicity, closed-form expression, and robust performance in balancing exploration and exploitation.
Formally, EI at a candidate point $x$ is defined as:
\begin{equation}
\text{EI}(x) = \mathbb{E}[\max(0, f(x) - f(x^+))],
\end{equation}
where $f$ is the objective function being optimized and $f(x^+)$ denotes the best observed objective value so far.

A shared limitation of conventional acquisition functions is that they are not task-specific: they are fixed and do not learn from prior experience, even when tasks share similar characteristics.
This missed opportunity limits their efficiency, especially in human-in-the-loop settings where evaluations are costly and rapid adaptation is required.
This motivates a shift toward learned acquisition functions, in which the acquisition function is learned from prior data.
The data is collected from similar users or, in our case, synthetic users, allowing it to generalize across users or tasks.

\paragraph{2. Deep reinforcement learning}

Neural Acquisition Function (NAF) is an approach that allows the optimizer (meta-BO) to learn a task-specific acquisition function using a neural network. 
Instead of relying on a fixed acquisition function like Expected Improvement (EI), NAF aims to learn a search strategy that is tailored to a type of optimization problems which shared similar characteristics.
A key challenge in learning such an acquisition function is the absence of ground truth labels. 
That is, there is no ``correct'' acquisition value to supervise the training. 
The predictions are inherently uncertain, noisy, and task-dependent.
To overcome this, NAF is trained using reinforcement learning (RL). 
The neural network generates acquisition function values across the design space. 
The design candidate with the highest score is selected for evaluation.
The outcome of this evaluation (i.e., the observed objective function value) is then converted into a reward signal: promising evaluation results yield positive rewards, while poor results generate negative feedback. 
Over time, the model learns to associate high acquisition values with promising candidates from its current status, refining its search policy.
Viewed from an RL perspective, meta-BO acts as an RL agent, and the neural acquisition function (NAF) serves as its policy model, determining which action (i.e., which design candidate) the agent should take in each iteration given the GP's status. 
This formulation enables task-level learning, where the acquisition policy improves through repeated interaction with synthetic training tasks.

Here, we briefly define the key terms of RL to better introduce NAF. 
\rev{For a more detailed introduction to reinforcement learning, we refer readers to \citet{sutton2018reinforcement} and  \citet{li2017deep}.}
RL is a machine learning paradigm that trains an agent to develop a policy for making decisions to maximize the reward signals within a given environment. 
\rev{The state that the agent observed at time step $t$ is usually denoted as $s_t \in \mathcal{S}$, where $\mathcal{S}$ is the state space. 
The agent's action at time step $t$ is denoted as $a_t$, which is determined by its policy $\pi$.
The action taken ($a_t$) will let the agent move onto the next state ($s_{t+1}$). 
Such a transition is also mostly based on a probability distribution, denoted as $p(s_{t+1} | s_{t}, a_{t})$. 
Accompanying the transmission from $s_t$ to $s_{t+1}$, the agent receives a reward signal, denoted as $r_{t}$, which is determined from a reward function ($r$).
We can further denote the reward received at $t$ as such, $r_{t} = r(s_{t}, a_{t}, s_{t+1})$.
The agent learns the policy via trial and error, aiming to maximize the cumulative discounted future reward based on its policy.}

Our \method follows \citet{volpp2019meta} to train the neural acquisition function using Proximal Policy Optimizations (PPO) \cite{schulman2017proximal}.
We frame each optimization task as an RL episode: the GP surrogate defines the state, the NAF outputs acquisition scores as the policy, and selecting a design corresponds to taking an action.
The observed objective value provides the reward.
Across many synthetic tasks, PPO updates the NAF to learn an acquisition policy that generalizes beyond any specific function.
Unlike prior work, our formulation incorporates human-centered requirements directly into the reward structure and training tasks, enabling the acquisition function to adapt to HCI scenarios. 
More details will be expanded in \autoref{sec:naf} and \autoref{tab:rl_mapping}.
The results showed that \method enables faster convergence than other methods, particularly showing statistically better performance than TAF in the second and third iterations.

\paragraph{3. Neural Acquisition Function (NAF)}
NAF keeps GP as the surrogate model of the task, allowing it to model the properties of the current optimization based on the real-time observations. 
NAF employs a neural network to generate the acquisition values.
In the original paper, NAF (the model itself) at timestep $t$ is $\alpha_{t,\theta}$, where $\theta$ indicates that it is parameterized by a neural network.
Similar to standard acquisition functions, NAF takes the information of the GP model as the input. 
A GP defines a continuous predictive function over the entire design space and can be queried at arbitrarily many points. 
In contrast, our NAF is a deep neural network that requires a fixed-dimensional input vector to summarize the whole GP's information.
Following \citet{volpp2019meta}, the design space $\mathcal{X}$ is discretized into a set of points ${x_s}$.
For each candidate $x$, NAF receives the GP's predicted mean $\mu_t(x)$ and standard deviation $\sigma_t(x)$ as input features.
Since the GP evolves with incoming observations, these features are dynamically updated at every iteration.
In the original implementation, the set of $x_s$ was selected dynamically at each step.
To simplify training and ensure stable input representation, we adopt a fixed set of $x_s$ throughout the optimization process.
Additionally, to help the model account for optimization progress and urgency, NAF includes a budget-aware input: the percentage of optimization completed, calculated as the current iteration index divided by the total allowed iterations.
This allows NAF to modulate its exploration–exploitation behavior over time.
The output of the neural network is a categorical distribution over the discretized candidates, where each probability corresponds to the predicted acquisition value of selecting a specific $x_s$.
During training, the next design candidate is sampled from this distribution, and the selected candidate is evaluated in the next iteration.
During deployment, NAF directly selects the design candidate that has the highest acquisition value (value generated from NAF).
We refer the readers to the original paper for more details about NAF \cite{volpp2019meta}.

\paragraph{4. Summary and Limitations of NAF}
NAF has a key advantage: it can be trained with unlimited data without incurring computational overhead at deployment.
This makes NAF particularly well-suited for our \conceptnospace, where synthetic users can be generated in virtually unlimited numbers via model sampling.
However, NAF has critical limitations when applied to real-world optimization tasks.
First, NAF is inherently designed to optimize for a single, fixed objective.
In practice, many interactive systems involve multi-objective trade-offs, such as balancing speed and accuracy in input tasks, or comfort and visibility in interface layout.   
NAF lacks the flexibility to adjust its optimization strategies when design objectives or user priorities shift.
Second, NAF, as a deep learning model, performs well only when test-time tasks fall within the distribution of training tasks. 
When applied to HILO, that is, NAF can only perform well when the real user's performance characteristics are highly aligned with the synthetic users' performance profile used during training.
NAF is likely to fail if real-world user performance deviates significantly from the synthetic users' performances (i.e., constitutes an out-of-distribution task), which is likely given that human behaviors are inherently noisy and diverse.

\begin{figure*}[h!]
\centering
\includegraphics[width=1\textwidth]{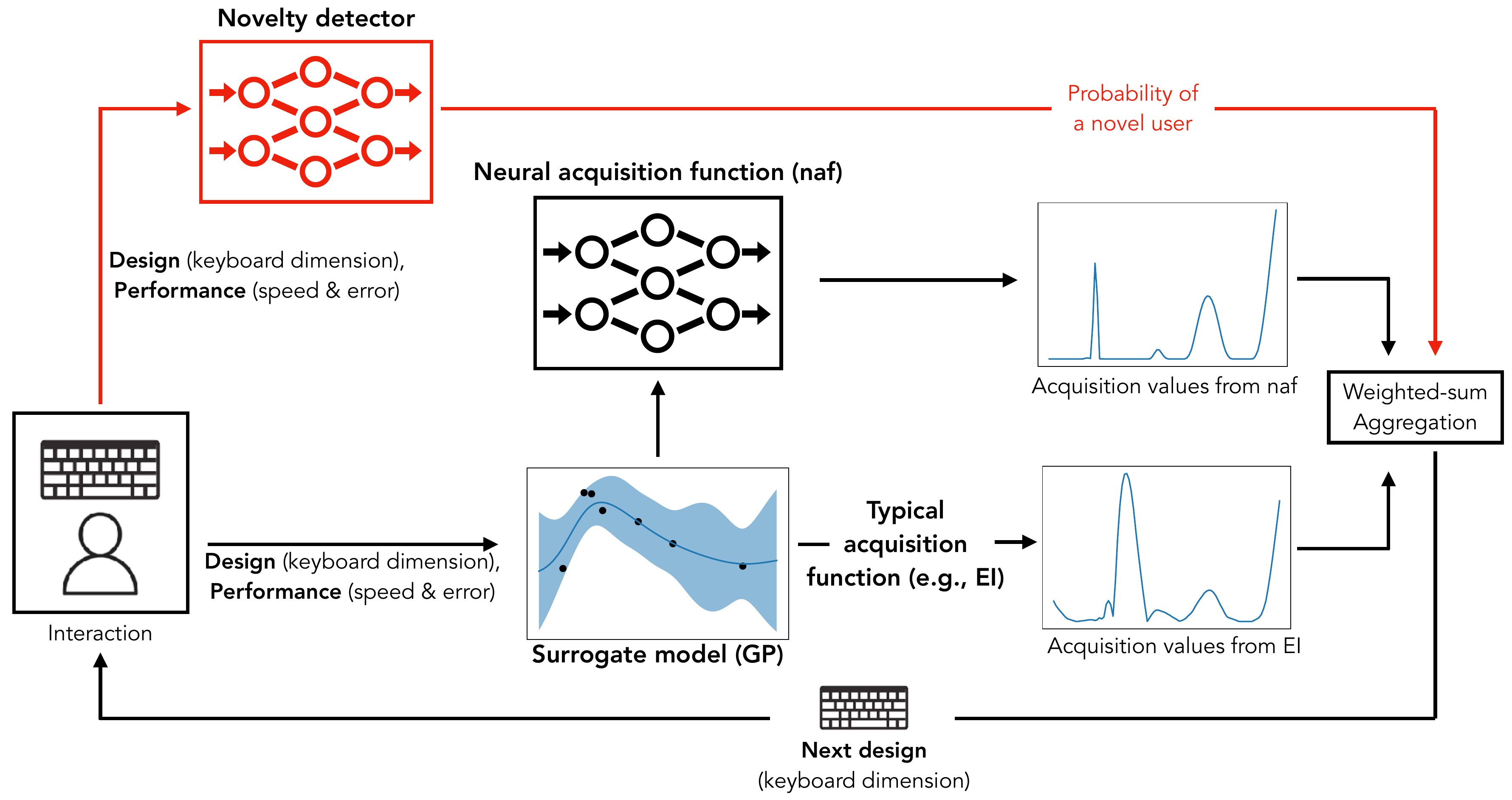}
  \caption{
  Our Neural Acquisition Function$^+$ (NAF$^+$) has four main components. Here, we illustrate how they work together using keyboard adaptation as an example case. The \textbf{surrogate model} (Gaussian Process regression) captures the properties of the target user. The information of GP will then allow the \textbf{neural acquisition function} (pre-trained by synthetic data) and the \textbf{typical acquisition function} to generate a set of acquisition values. A \textbf{novelty detector} estimates how likely the new user is different from our training dataset based on the current observations; this information is then used to condition the aggregation of two acquisition functions. Finally, NAF$^+$ suggests a design that is most likely to yield optimal user performance. 
  }
 ~\label{fig:naf}
\end{figure*}

\subsection{Overview of \method}

To address the aforementioned challenges, we introduce \methodnospace, a new approach that builds upon NAF but incorporates two critical enhancements: the ability to dynamically adjust to multi-objective formulations, and fallback mechanisms for handling out-of-distribution users.
These enhancements make \method more flexible, generalizable, and better-suited for real-world human-in-the-loop applications.
We assume that the target application involves \textit{multiple design objectives}, which are eventually aggregated into a single scalar objective via a weighted-sum formulation; this is an approach commonly found in HCI work~\cite{10.1145/3613904.3642071, bailly2013menuoptimizer}.
Our contribution lies in training \method such that it can support dynamic setting of objective weights during deployment without requiring any retraining, which allows the optimizer to respond to varying user needs or preferences on the fly.
We also assume that while novel users (whose behavior significantly deviates from synthetic users) may appear during deployment, they can still benefit from standard BO mechanisms (e.g., exploration via typical acquisition functions).
Therefore, \method integrates a \textit{fallback mechanism} to maintain robustness when facing novel, unseen users. 

As illustrated in \autoref{fig:naf}, \method has four key components to jointly achieve these enhancements: (1) \textbf{GP-based surrogate model (GP)}: This component models the objective function based on the user's observed performance data. 
(2) \textbf{Neural acquisition function (\texttt{naf})}: This deep neural network replaces hand-crafted acquisition functions by learning a task-specific policy. Unlike the original NAF, our model is explicitly trained to condition on \textit{objective weights}, enabling it to adjust its search strategy according to the prioritized trade-offs during deployment. To distinguish it from the original method, we refer to our version in lowercase: \texttt{naf}.
Two more elements are introduced to mitigate the novel user challenge: (3) \textbf{Expected Improvement (EI):} An acquisition function used as a \textit{fallback}. When the neural model performs poorly (e.g., due to encountering a novel user), EI provides a stable and exploration-oriented alternative. The acquisition values produced by \texttt{naf} and EI are combined via a weighted sum.
(4) \textbf{Novelty detector:} This module estimates how novel the current user is with respect to the training distribution. Based on this assessment, it dynamically adjusts the weight between the neural acquisition score (\texttt{naf}) and the EI score. As novelty increases, more weight is placed on EI to ensure robust performance.

Note that two of the four elements (the GP and the EI) follow standard BO, and the \texttt{naf} component is adopted from prior work \cite{volpp2019meta}.
Our contribution lies in integrating these components and adapting them to the unique challenges of human-centered optimization, including dynamically weighted objectives and out-of-distribution users.
We expand the details of each component below.

\subsection{Element 1: GP-based Surrogate model (GP)}

The surrogate model reflects the ongoing user's performance. We follow the typical BO framework and employ a standard GP. Please refer to \autoref{sec:pre} for a more detailed introduction.

\subsection{Element 2: Neural Acquisition Function (\texttt{naf})}
\label{sec:naf}

The \texttt{naf} model in \method is a neural network trained to generate acquisition values for a specific interaction scenario, learned through interactions with synthetic users.  
We denote the \texttt{naf} model as $\alpha_{t, \theta}$, where $t$ indicates the current timestep and $\theta$ corresponds to the neural network parameters.
Our training process follows the general framework introduced in the original NAF paper~\cite{volpp2019meta}, with key extensions tailored to the requirements of HCI tasks.

First, we discretize the design space $\mathcal{X}$ into a fixed grid of candidates $\{x_s\}$.  
For each candidate $x_i$, where $i$ indexes the grid, the surrogate model (a Gaussian Process) provides a predicted mean $\mu_t(x_i)$ and standard deviation $\sigma_t(x_i)$ at optimization iteration $t$.  
These values form the core input features to \texttt{naf}, representing the model's current belief about the design space.
To enable dynamic adaptation over time, we include a budget-awareness signal: the proportion of the optimization process completed, calculated as $t/T$, where $T$ is the total optimization budget.  
This allows \texttt{naf} to shift its strategy from exploration in early iterations to exploitation in later stages.

A key innovation in \method is its ability to support dynamic multi-objective optimization during deployment.  
To enable this, we append $N$ additional input values to the input vector, where $N$ is the number of design objectives.  
These values represent the weights assigned to each objective under the current trade-off configuration (e.g., $w_1, w_2, \dots, w_N$).  
For example, in a two-objective task with equal weights $[0.5, 0.5]$, the input vector to \texttt{naf} will include these two values at the end, allowing the model to adapt its acquisition strategy accordingly.
The full input to \texttt{naf} at iteration $t$ is thus:
\begin{equation}
[\mu_t(x_1), \sigma_t(x_1), \dots, \mu_t(x_s), \sigma_t(x_s), t/T, w_1, w_2, \dots, w_N].
\end{equation}

\noindent
The output of \texttt{naf} is a categorical distribution over the $s$ candidate points, denoted as $\text{CAT}[\text{nv}_{x_1}, \text{nv}_{x_2}, \dots, \text{nv}_{x_s}]$. 
Essentially, each $\text{nv}_{x_i}$ value represents the \texttt{naf}-generated acquisition value at a given design candidate $x_i$ across the full grid $x_s$.
\autoref{tab:rl_mapping} summarizes the RL formulation used in our meta-BO training pipeline.
During training, a candidate design is sampled from this distribution to encourage exploration.  
During deployment, however, the design with the highest predicted acquisition probability is selected deterministically to maximize performance.

\begin{table*}[h]
\centering
\begin{tabular}{|l|l|}
\hline
\textbf{RL Component}     & \textbf{\method Equivalent} \\
\hline
Policy $\pi_\theta$       & Neural acquisition function (\texttt{naf}) $\alpha_{t,\theta}$ \\
Episode                  & Optimization episode with a synthetic user \\
Episode length $T$        & Optimization budget $T$ \\
State $s_t$              & $[\mu_t(x_s), \sigma_t(x_s), t/T, w_1, \dots, w_N]$ \\
Action $a_t$             & Selecting a design candidate $x_t \in \mathcal{X}$ \\
Reward $r_t$             & $- (f_{\text{max}} - f(x_t))$ (negative regret) \\
Transition $p(s_{t+1} | s_t, a_t)$ & Noisy observation of $f(x_t)$, GP update \\
\hline
\end{tabular}
\caption{The RL formulation and mapping of \methodnospace.}
\label{tab:rl_mapping}
\end{table*}

\subsection{Element 3: Expected Improvement}

A typical EI serves as the fallback acquisition strategy, with details provided in \autoref{sec:pre}.
We apply EI to each discretized candidate $x_i \in \{x_s\}$ in the design space, producing a parallel set of acquisition scores $\text{EI}(x_i)$.
These scores are then combined with the values generated by the neural acquisition function (\texttt{naf}) to form the final acquisition values used for decision-making.
The weighting between \texttt{naf} and EI is dynamically adjusted based on a novelty-aware mechanism, described in the following section, ensuring robustness when encountering out-of-distribution users.

\subsection{Element 4: Novelty Detector}

The novelty detector in \method is a probabilistic module that estimates how likely a new user is to be out-of-distribution (i.e., significantly different from all previously seen synthetic users). 
Its primary role is to dynamically determine the weighting between the neural acquisition function (\texttt{naf}) and the fallback strategy (EI) when selecting the next design candidate.

At its core, the novelty detector is implemented as a Bayesian Neural Network (BNN), trained using the full set of $(x, y)$ pairs --- design candidates and their corresponding objective function values --- collected from synthetic users during pre-training. 
This training allows the BNN to approximate the population-level distribution of performance across diverse user characteristics, effectively modeling expected outcomes under a wide range of known conditions.

During deployment, when a new observation $(x, y)$ is collected from a real user, the BNN outputs a posterior predictive distribution for $x$, characterized by a mean $\mu_t(x)$ and standard deviation $\sigma_t(x)$. 
Using this, we compute the standardized z-score:

\begin{equation}
z = \frac{y - \mu_t(x)}{\sigma_t(x)}
\end{equation}

We then convert the z-score into a two-tailed p-value using the cumulative distribution function $\Phi(\cdot)$ of the standard normal distribution:
$p = 2 \cdot \left(1 - \Phi(|z|)\right)$. 
The p-value quantifies how surprising the real user's performance is under the model trained on synthetic users. 
A low p-value (e.g., close to 0) indicates high novelty and suggests the user is likely out-of-distribution, whereas a high p-value indicates the user is well represented in the training data.

Over time, we accumulate multiple $(x, y)$ pairs from the same user and compute a running average of the p-values, denoted $\bar{p}$. 
We use this average to determine the contribution of EI in the acquisition strategy.
Specifically, if $\bar{p}$ falls below a pre-defined novelty threshold $\tau$, we consider the user to be novel. 
The final acquisition score $A(x)$ used to select the next design candidate is computed as a weighted combination of the \texttt{naf} and EI scores:

\begin{equation}
\label{eq:weighted-sum}
A(x) = \lambda_{\text{EI}} \cdot \text{EI}(x) + (1 - \lambda_{\text{EI}}) \cdot \texttt{naf}(x)
\end{equation}
where the EI weight $\lambda_{\text{EI}}$ is computed as:

\begin{equation}
\lambda_{\text{EI}} = 
\begin{cases}
1 & \text{if } \bar{p} = 0 \\
0 & \text{if } \bar{p} \geq \tau \\
\frac{\tau - \bar{p}}{\tau} & \text{otherwise}
\end{cases}
\end{equation}

This weighting mechanism ensures a smooth transition: when the user appears entirely novel ($\bar{p} = 0$), full weight is given to EI; when the user aligns with the training distribution ($\bar{p} \geq \tau$), EI is ignored; otherwise, the influence of EI scales linearly based on the level of novelty.


\subsection{Training and Deployment}

\texttt{naf} is trained with a large number of synthetic users.
Then, we obtain a large collection of design-performance pairs $(x, y)$ representing population-level behavior, which are used to train the novelty detector.
During \emph{deployment}, the novelty detector assesses the novelty of the user and adjusts the weight assigned to EI, as described in \autoref{eq:weighted-sum}.
We set a minimum iteration threshold $K=3$, so that the weighted-sum mechanism for acquisition values is activated from the 4th iterations, ensuring that \texttt{naf} maintains full control during the initial 3 iterations.

\subsection{Synthetic tests}

Before conducting the user study, we designed two synthetic evaluations to systematically validate \method. 
These tests serve as controlled environments to probe capabilities that are difficult to isolate in live studies. 
Specifically, we focused on three goals: (\textbf{G1}) benchmarking against established baselines: standard BO and Transfer Acquisition Function, which is an established meta-BO method used in prior HCI works \cite{wistuba2018scalable, 10.1145/3613904.3642071};
(\textbf{G2}) verifying that \method can adapt to dynamic trade-offs between multiple objectives via explicit weight inputs;
(\textbf{G3}) evaluating robustness when encountering users whose behavior falls outside the training distribution. 
To cover both abstraction and realism, we used two tasks: a controlled double-Sphere benchmark (Test~1) for clean comparisons, 
and a semi-realistic typing simulation (Test~2) grounded in Fitts' law and touch error models on a touchscreen to mirror our target VR keyboard task in the user study.

\paragraph{Key findings.} 
Across both tests, \method consistently demonstrated advantages over baselines. 
For \textbf{G1 (baseline comparison)}, it converged faster than standard BO and TAF, particularly in the early iterations where efficiency is most critical. 
For \textbf{G2 (dynamic objective weighting)}, conditioning on explicit weight inputs enabled \method to adapt its search strategy more effectively than the ablated variant without weights, confirming its ability to handle shifting priorities in multi-objective optimization. 
For \textbf{G3 (robustness to novel users)}, the novelty-aware fallback proved essential: when synthetic users were sampled outside the training distribution, \method recovered and improved steadily, while the no-fallback variant stalled and BO/TAF required more iterations to catch up. 
Together, these results establish that \method combines sample efficiency, flexibility, and robustness, making it well-suited for real-world HILO where the weights on multiple objective can be adjusted dynamically and users can vary dramatically.
The results of the synthetic tests are presented in the Appendix.

\section{User Study: \method for Mid-Air Keyboard Adaptation}

We conducted a user study to evaluate the effectiveness of \method in a classic HCI task: mid-air typing in virtual reality; the interaction is shown in \autoref{fig:study_interface}.
Participants completed ten iterations of typing (each consisting of one to two sentences) with adaptive keyboards with varying key height and width, generated by three methods. 
Subsections~5.1--5.4 describe the study design, including the experimental conditions (methods), the user models used to train \method, the objective function, and the study apparatus. 
The study was conducted in two phases. 
Phase~1 (Subsection~5.5) provided data to pre-train two optimization methods, TAF and \methodnospace. 
Subsection~5.6 details the implementation and training procedure of \method. 
Phase~2, the final evaluation, is described in Subsections~5.7 (study setup) and~5.8 (results). 
Finally, Subsection~5.9 presents our discussion of the study findings.

\subsection{Overall Study Design and Conditions}

We compared three optimization methods: our proposed \methodnospace, Transfer Acquisition Function (TAF) \cite{wistuba2018scalable,10.1145/3613904.3642071}, and Continual Bayesian Optimization (ConBO) \cite{liao2025continual}. 
Both TAF and ConBO represent state-of-the-art baselines. 
TAF adopts a meta-learning approach, where prior user data from complete HILO processes is leveraged to construct transferable models for new users. 
ConBO instead follows a continual learning approach: each user's data directly serves as prior knowledge for subsequent users, without explicitly distinguishing between prior and target users.  

To mitigate fatigue and learning effects, we constrained the study duration to one hour. 
Under this constraint, we limited the evaluation to three adaptation methods. 
Prior investigations have shown that standard Bayesian Optimization typically underperforms compared to approaches augmented with prior data \cite{doi:10.1177/26339137241241313,liao2025continual}, so we excluded it as a baseline. 
Similarly, prior work demonstrated that manual calibration achieves performance comparable to prior-data–based optimization \cite{liao2025continual,10.1145/3613904.3642071}, though in a substantially different setting where participants tuned the interface in advance. 
We therefore also excluded manual calibration from our study conditions.  

Among the three selected methods, both \method and TAF require pre-training, and the study was therefore organized into two phases. 
Phase~1 collected typing data to pre-train both methods, following protocols established in prior work~\cite{10.1145/2984511.2984546,10.1145/3613904.3642071}. 
For \method, this data was used to fit the parameters of existing analytical typing models (Fitts' Law and the DGD model) to our specific interaction context. 
While these models are well-established, no prior work has provided fitted parameters for the mid-air keyboard with direct touch; future work could bypass this step by reusing our fitted distributions. 
Also, other applications can potentially bypass this step and directly train \method if reliable model parameters have already been estimated, such as touchscreen interactions where typing errors and movement time are well studied.
For TAF, the same Phase~1 data was required to construct Gaussian Process (GP) models of prior users. 
Using a shared dataset for both methods ensures fairness: since TAF necessarily requires empirical data, we leverage the same data to calibrate \methodnospace.

\subsection{Typing Simulation for Training \method}
\label{sec:user_typing_model}

To train \methodnospace, we implemented a generative simulation of sentence typing that models the motor execution and touch uncertainty of a human typist. 
The simulator combines two established models of touchscreen interaction: Fitts' Law for movement time estimation~\cite{mackenzie1992extending} and the Dual Gaussian Distribution (DGD) model for touch accuracy~\cite{10.1145/2470654.2466180,10.1145/2984511.2984546,10.1145/2501988.2502058}.  

Each simulated user is parameterized by a 6-dimensional vector 
$\theta = [a, b, \alpha_x, \sigma_{\alpha_x}, \alpha_y, \sigma_{\alpha_y}]$, 
which captures their motor and perceptual characteristics. 
The parameters $a$ and $b$ govern movement speed, while $\alpha$ and $\sigma$ terms control touch precision and variability in the horizontal and vertical directions.  

\paragraph{Movement time.}  
The time to move between successive keystrokes is modeled using the classical Fitts' Law equation:  

\begin{equation}
MT = a + b \cdot \log_2\!\left(\frac{D}{W} + 1\right),
\end{equation}

\noindent
where $D$ is the Euclidean distance between the current and target key centers, and $W$ is the smaller dimension of the key.  

\paragraph{Touch distribution.}  
Upon reaching a target key, the simulator samples a touch point from a 2D Gaussian distribution centered on the intended key:  

\begin{equation}
\Sigma =
\begin{bmatrix}
\alpha_x \cdot W^2 + \sigma_{\alpha_x}^2 & 0 \\
0 & \alpha_y \cdot H^2 + \sigma_{\alpha_y}^2
\end{bmatrix}.
\end{equation}

\noindent
This covariance formulation reflects the DGD model, where touch precision is influenced by both absolute noise ($\sigma$) and a size-dependent term ($\alpha$). 
Unlike prior work that analytically integrates over the distribution~\cite{10.1145/2984511.2984546}, we simulate keystrokes via Monte Carlo rollouts: each touch sample is checked against the target key boundaries, and marked as correct or erroneous.  

\paragraph{Sentence typing.}  
Test sentences are drawn from the Enron Mobile Email Dataset\footnote{\url{https://www.keithv.com/software/enronmobile/}}, with special characters removed and concatenated to a length at least 26 characters. 
Typing proceeds sequentially: the simulator begins at the spacebar, moves to each target key, samples a noisy landing point, and records both movement time and correctness. 
Additional Gaussian noise ($\sigma = 0.15$ s) is added to capture motor variability.  
This procedure yields average movement time and error rate for a given keyboard layout.  
These settings remain identical in both our simulation experiments (Appendix~B) and the user study.

\begin{figure*}[h!]
\centering
\includegraphics[width=1\textwidth]{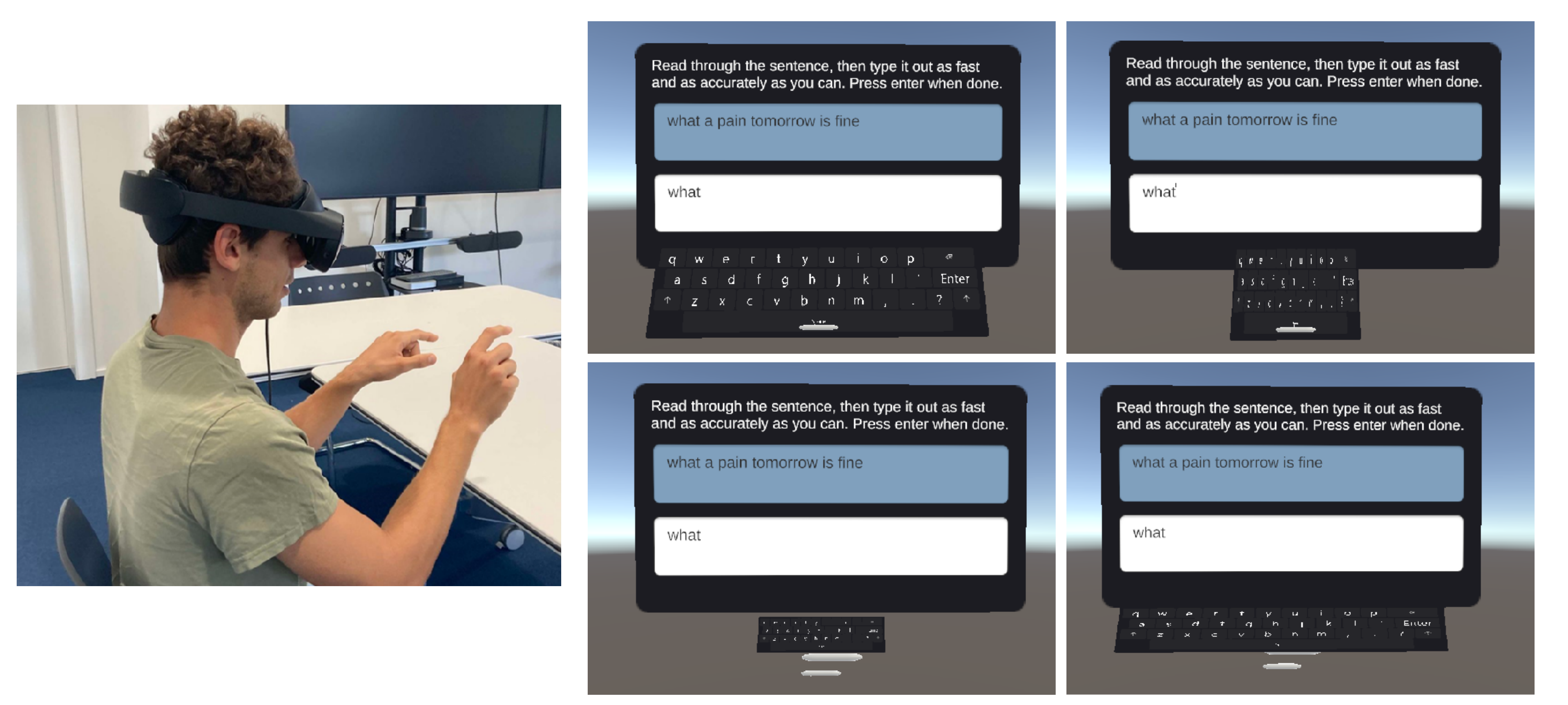}
  \caption{
   Our user study aims to personalize the mid-air keyboard's key height and width using three methods: our \method, Transfer Acquisition Function (TAF), and Continual Bayesian optimization (ConBO). TAF and ConBO are both existing BO approaches for achieving efficient HILO with prior experience. Here we present some possible keyboard variants within the design parameter range.
  }
 ~\label{fig:study_interface}
\end{figure*}

\subsection{Objective Function and Task Setting}
\label{sec:objective_function}

Our goal is to optimize the key's height and width of mid-air keyboards. 
The simulator outputs two primary metrics: typing speed (words per minute, WPM) and error rate. 
To combine them into a single scalar measure, we normalize each into unitless functions of \textit{Speed} and \textit{Accuracy}.  

Through a 3-person pilot test, we determined that typing speed typically ranged between 5~WPM (worst) and 22~WPM (best). 
We therefore linearly normalized this range such that 5~WPM maps to 0 and 22~WPM maps to 1.  
Error rate, computed via Levenshtein distance, was observed to range between 0\% (best) and 30\% (worst). 
We similarly normalized this range so that 0\% error maps to 1 and 30\% to 0.  
While a 30\% error rate may appear higher than in prior work, this upper bound is necessary in our adaptation setting, where small keyboards can lead to frequent errors.  

Finally, speed and accuracy are combined in a weighted objective function:  

\begin{equation}
\text{Objective} = w_{\text{speed}} \cdot \text{Speed} \;+\; w_{\text{accuracy}} \cdot \text{Accuracy},
\label{eq:objective}
\end{equation}

\noindent
where we set $w_{\text{speed}} = 0.7$ and $w_{\text{accuracy}} = 0.3$ for the two phases in the user study. 
Other studies may adjust the weights to reflect different priorities.  

In short, this setup captures the fundamental trade-off in mid-air typing: larger keys improve accuracy but increase movement times, while smaller keys accelerate typing but risk higher error rates. 
We therefore restrict the design space to keyboard widths and heights between 20\,mm and 40\,mm, and optimize within this range to maximize the objective in \autoref{eq:objective}.

\subsection{Study Apparatus and Implementation}

The virtual keyboard and typing experiment were implemented in Unity and executed on a Windows~10 desktop with an NVIDIA GeForce RTX~3050 GPU. 
We built the keyboard on top of the MRTK keyboard framework\footnote{\url{https://github.com/microsoft/MixedRealityToolkit-Unity}}, which we customized to support adaptive resizing of key height and width. 
The study was conducted on a Meta Quest Pro headset, as shown in \autoref{fig:study_interface}.
Target sentences were sampled from the Enron Mobile Email Dataset, with special characters removed and multiple sentences concatenated to ensure a fixed length of 26 characters (excluding the space key) per iteration.

\subsection{Phase 1: Gathering Data for \method and TAF}

The goal of Phase~1 was to collect training data to acquire typing models for \method and Gaussian Process (GP) models for TAF.  

\subsubsection{Procedure}

Five participants (2 female, 3 male; age 27--32) were recruited for Phase~1. 
Participants were instructed to type the presented sentences as quickly and accurately as possible on the mid-air keyboard. 
Before the main session, each participant completed 10 practice rounds with randomly varying keyboard dimensions to familiarize themselves with the adaptive setting.  
During the experiment, the keyboard remained at a fixed position (30\,cm in front of the participant and tilted at $-20^{\circ}$ relative to the horizontal plane), while only its width and height adapted.  
Each participant then typed 20 sentences, making the session last approximately 30 minutes, including practice.  

To acquire data suitable for both TAF and \methodnospace, we adopted a hybrid procedure. 
In the first five iterations, keyboard dimensions were selected from evenly spaced values across the design space ($[20,25,30,35,40]$\,mm) for both width and height, presented in randomized order. 
These evenly spaced settings ensure sufficient coverage for fitting typing error models (\autoref{sec:user_typing_model}). 
For the remaining 15 iterations, we used Bayesian optimization: the weighted objective (Speed + Accuracy) from prior trials was fitted to a surrogate model, and the acquisition function suggested new candidate designs.  
The full set of 20 iterations thus provided both uniformly sampled data for model fitting and adaptively sampled data for training GP priors in TAF.  

Within each iteration, 26 key presses were recorded. 
We omitted the first six presses as a warm-up, leaving 20 effective key presses per iteration.  
Participants were instructed not to delete or correct typos. 
In cases where a participant attempted correction (e.g., typing ``a'' instead of ``s,'' then pressing ``s'' again), only the first (incorrect) press was retained for analysis.

\subsubsection{Results}

From the collected data, we constructed five GP models to serve as prior models for TAF.  
In parallel, we applied simple linear regression to approximate parameter values of the typing models.  
For each participant, individual parameter values were estimated and then summarized across participants as means and standard deviations.  

For Fitts' Law, the parameters were $a = 0.164 \pm 0.0352$\,s and $b = 0.39 \pm 0.171$\,s.  
For the error models, we obtained $\alpha_x = 0.0148 \pm 0.0011$, $\sigma^2_x = 15.52 \pm 2.093$, $\alpha_y = 0.0133 \pm 0.0011$, and $\sigma^2_y = 15.93 \pm 1.46$.  
These values provide empirical ranges for simulated users and serve as priors for model-based optimization in Phase~2.

\subsection{Implementing and Training \method} 

Before Phase~2 of the user study, we prepared \method by generating synthetic users, and implementing and training its components.  

\subsubsection{Creating Synthetic Users}

Based on the parameter distributions estimated in Phase~1, we generated synthetic users by sampling from normal distributions over the six parameters 
($a$, $b$, $\alpha_x$, $\sigma_{\alpha_x}$, $\alpha_y$, $\sigma_{\alpha_y}$). 
Each sampled parameter set corresponds to a unique synthetic user.  
Given a keyboard configuration, a synthetic user simulates typing behavior and outputs Speed and Accuracy, which are combined via the objective function (\autoref{eq:objective}).  
This approach enables large-scale training of \methodnospace without additional human data collection.  

\subsubsection{Implementation of \method}

\method consists of three components: the \texttt{naf} policy network, a BNN-based novelty detector, and a GP model.  
The \texttt{naf} network is a six-layer fully connected neural network, with 256 units per hidden layer and Tanh activations.  
The novelty detector is implemented as a two-layer fully connected Bayesian neural network (BNN) with 256 units per layer, using dropout ($p = 0.2$) to approximate weight uncertainty.  
At inference time, multiple stochastic forward passes are performed to estimate predictive mean and variance, which are converted into z-scores and novelty p-values.  
The GP model is implemented using the \texttt{botorch} library as a single-task Gaussian Process\footnote{\url{https://botorch.org/docs/models/}}.

\subsubsection{Training of \method}

The \texttt{naf} network is trained using Proximal Policy Optimization (PPO)~\cite{schulman2017proximal}.  
The actor and critic share the same architecture, with learning rates of $0.0003$ and $0.001$, respectively.  
PPO is run for $K = 10$ epochs with a clipping parameter of 0.15.  
Although the user study employs fixed weights ($w_{\text{speed}} = 0.7$, $w_{\text{accuracy}} = 0.3$), we train \texttt{naf} to generalize across different weightings by randomly sampling $(w_{\text{speed}}, w_{\text{accuracy}})$ during training.  
Synthetic users are dynamically generated throughout training, and the process continues until 80,000 steps are reached.  

For the novelty detector, we consider 11 weight combinations by setting $w_{\text{speed}} \in \{0, 0.1, 0.2, \ldots, 1.0\}$ and $w_{\text{accuracy}} = 1 - w_{\text{speed}}$.  
For each weight setting, we sample 20 synthetic users and evaluate 100 keyboard layouts (grid search), yielding 2,000 performance profiles per weight condition.  
These datasets are then used to train BNN models for novelty detection, each trained for 2,000 epochs.  

\subsection{Phase 2: Evaluation of Different Methods}

In Phase~2, we evaluated three optimization methods for human-in-the-loop mid-air keyboard adaptation: our proposed \methodnospace, Transfer Acquisition Function (TAF), and Continual Bayesian Optimization (ConBO).  

\subsubsection{Study Design and Procedure}

The study employed a within-subject design with 12 participants (5 female, age 24--33).  
Each participant completed three blocks, one per optimization method, with order counterbalanced using a Latin square.  
In each block, participants typed 10 sentences (iterations).
Before any block, the participants had a short practice session of 10 sentences where the keyboard dimension changed randomly.  
Between blocks, a 5-minute break was provided to reduce fatigue.  
The entire session lasted less than 60 minutes per participant.  
Finally, the participants were asked to fill in the raw NASA-TLX questionnaire followed by an open-ended interview.

\subsubsection{Methods}

\paragraph{\methodnospace.}  
We described the implementation of \method in Section~5.6.  
The novelty detector automatically selected the correct objective weight combination ($w_{\text{speed}} = 0.7$, $w_{\text{accuracy}} = 0.3$).  
We set the novelty detector threshold $\tau = 0.1$. This choice aligns with common statistical practice, where a p-value below $0.1$ is often interpreted as an indicator of outliers or model mismatch. 
Because our novelty estimate is based on the running average of p-values across iterations, a slightly more permissive threshold also ensures that the system can detect meaningful deviations early. 
This is also the value we used in the synthetic tests, shown to be an effective setting. 
Importantly, the novelty detector begins at the fourth iteration. 
In the initial three iterations, EI weights are set to be $0$, and the weights assigned on \texttt{naf} is set to be $1$.

\paragraph{Transfer Acquisition Function (TAF)}  
TAF relies on a set of prior models—in our case, five GP models trained on the Phase~1 data.  
During deployment, acquisition function values are aggregated across these prior models using adaptive weights.  
The weights assigned to prior users decay according to two hyperparameters ($\alpha_1, \alpha_2$), so that the influence of the new user's model increases over time, allowing the optimization to converge toward user-specific preferences.  
We followed the parameterization of prior work~\cite{10.1145/3613904.3642071}, setting $\alpha_1 = 4$ and $\alpha_2 = 0.2$.  

\paragraph{Continual Bayesian Optimization (ConBO)}  
ConBO relies on a Bayesian neural network (BNN) population model trained on all previously observed users.  
Unlike TAF, ConBO does not reuse Phase~1 data; instead, the 12 participants in Phase~2 sequentially contribute training data to the population model.  
For example, data from Participant~1 updates the model, which then serves as prior knowledge for Participant~2, and so forth.  
As in TAF, the final acquisition function balances the population model with the current user's model, assigning increasing weight to the latter to ensure personalized optimization.  
We largely followed the implementation of the original ConBO paper~\cite{liao2025continual}, with the same hyperparameters ($r_0 = 6$, $d_r = 2$).  
To align with TAF, we set the decay parameters to $\alpha_1 = 4$ and $\alpha_2 = 0.2$.

\subsubsection{Results}

\begin{figure*}[t]
\centering
\includegraphics[width=1\textwidth]{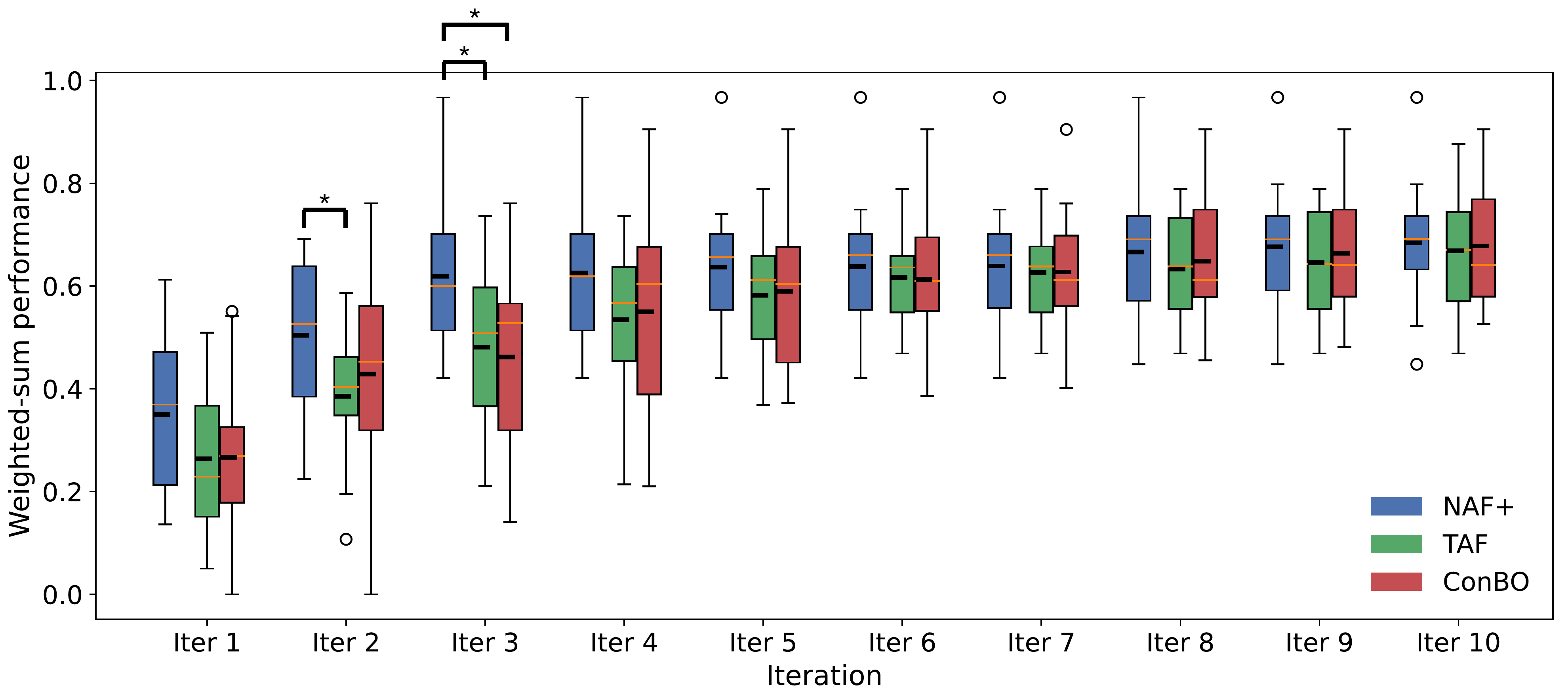}
  \caption{
  The weighted-sum performance achieved by different conditions (optimization methods) at each iteration. This plot presents the running best performance (i.e., the best performance of each participant achieved up to this iteration), which is a common way of analyzing iterative optimization tasks. The * sign indicates significant differences ($p<0.05$). The thicker, black line in each box shows the mean value, and the thin, orange line indicates the median value.
  The results showed that our \method enables faster convergence than other approaches. In particular, \method achieved statistically better performance than TAF in the second and third iterations, and outperformed ConBO in the third iteration.
  }
 ~\label{fig:study_result}
\end{figure*}

\begin{figure*}[t]
\centering
\includegraphics[width=1\textwidth]{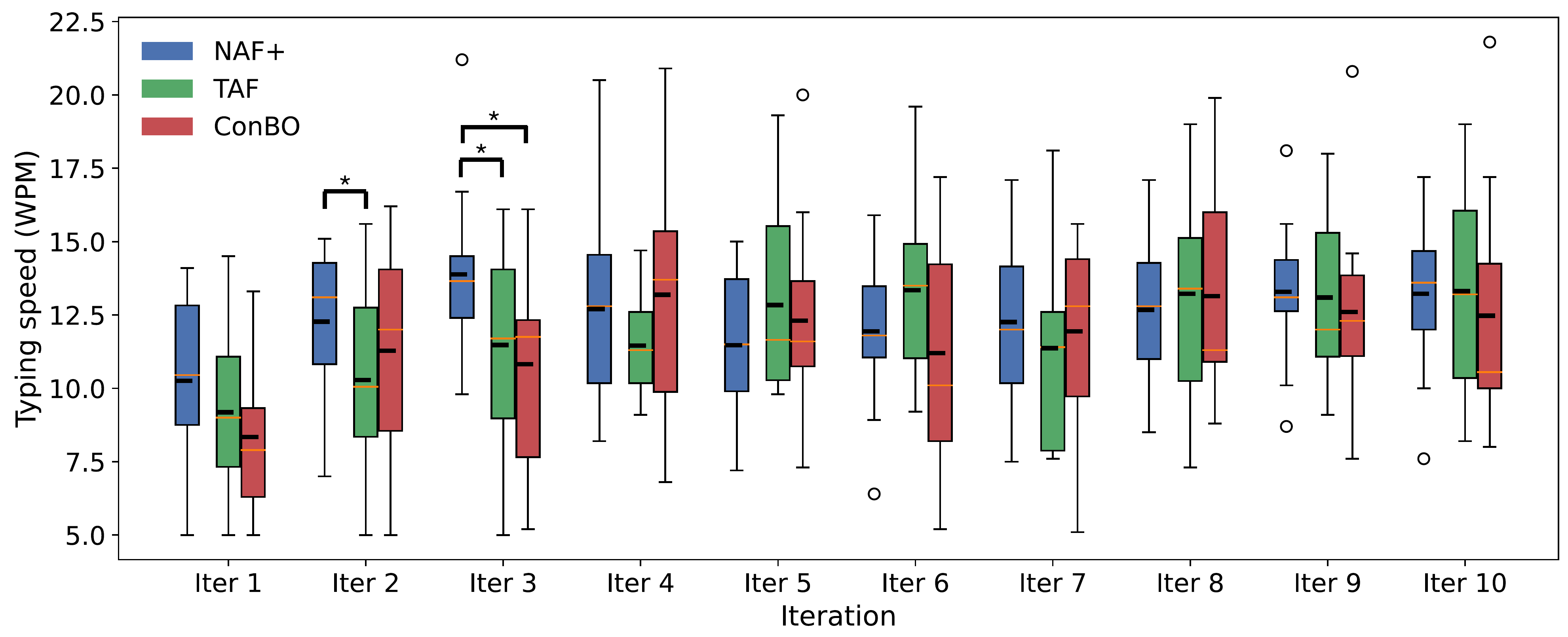}
  \caption{
  The typing speed at each iteration, which is then contributed to the calculation of the final objective function as described in \autoref{sec:objective_function}. The * sign indicates significant differences ($p<0.05$). The thicker, black line in each box shows the mean value, and the thin, orange line indicates the median value. 
  The results showed that \method enabled faster typing speed than TAF in the second and third iterations, and faster than ConBO in the third iteration.
  }
 ~\label{fig:wpm}
\end{figure*}

\begin{figure*}[t]
\centering
\includegraphics[width=1\textwidth]{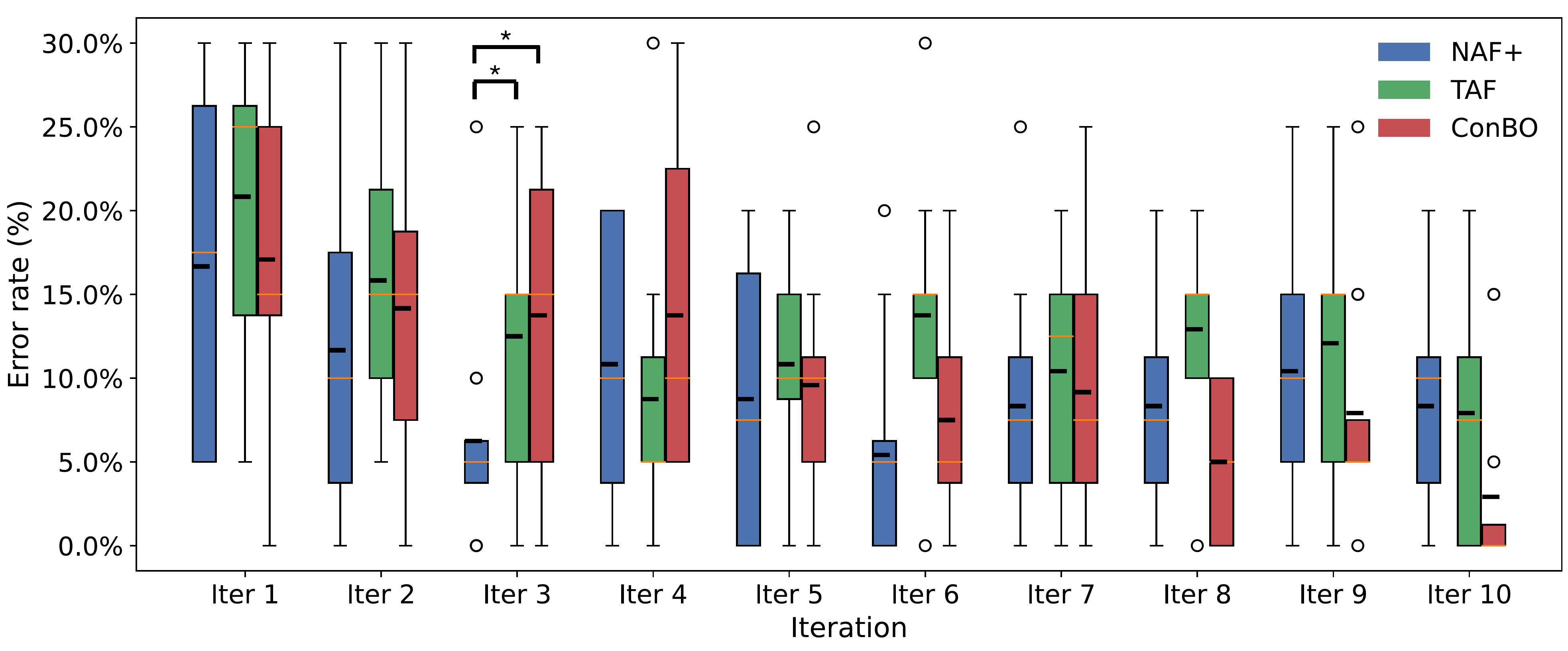}
  \caption{
  The error rate at each iteration, which is then contributed to the calculation of the final objective function as described in \autoref{sec:objective_function}. 
  The * sign indicates significant differences ($p<0.05$). The thicker, black line in each box shows the mean value, and the thin, orange line indicates the median value.
  The results showed that \method led to lower error rates than the other methods in the third iteration.
  }
 ~\label{fig:error_rate}
\end{figure*}

To evaluate optimization performance across iterations, we computed the objective function achieved in each iteration and method for every participant.
The final objective function values for each condition (optimization method) at each iteration are shown in \autoref{fig:study_result}. Note that we show and analyze the running best objective function values (i.e., the best performance achieved up to this iteration), which is a common practice for analyzing iterative optimization tasks in both machine learning and HCI fields. 
The raw typing speed (\autoref{fig:wpm}) and the raw error rate (\autoref{fig:error_rate}) at each iteration are also shown, which jointly contribute to the final objective function.

We conducted a two-way repeated-measures ANOVA with factors \textit{Iteration} (10 levels) and \textit{Method} (\method, TAF, ConBO). 
There is a strong main effect of Iteration, $F(9, 99) = 27.16$, $p < .001$, indicating that participants' performance, overall, improved over time. 
There was no significant main effect of Method, $F(2, 22) = 1.25$, $p = .307$, nor a significant Iteration $\times$ Method interaction, $F(18, 198) = 1.27$, $p = .211$. 
This indicates there is no method that is constantly better than the other throughout all iterations, which is aligned with the observation where three methods all converge toward the same optimal performance in later iterations (e.g., Iteration 6 onward). 
To further investigate whether different methods yield differences at different iterations, especially early iterations, we conduct one-way repeated-measures ANOVAs on each iteration separately. 
Significant effects of Method were found at Iteration~2, $F(2, 22) = 3.74$, $p = .040$, and Iteration~3, $F(2, 22) = 8.02$, $p = .002$, but not at any other iteration (all $p > .0.05$). 
Post-hoc pairwise comparisons with Bonferroni correction showed that at Iteration~2, \method outperformed TAF, $t(11) = 3.49$, $p_{\text{bonf}} = .015$, with no significant differences involving ConBO. 
At Iteration~3, \method outperformed both TAF, $t(11) = 3.52$, $p_{\text{bonf}} = .014$, and ConBO, $t(11) = -3.41$, $p_{\text{bonf}} = .018$, while TAF and ConBO did not differ significantly.  
The results show that while participants improved steadily across iterations regardless of method and all methods ultimately converge to comparable performances, \method provided a performance advantage over TAF in earlier iterations (at Iterations~2--3) and over ConBO (at Iteration~3 only).  

\autoref{fig:wpm} and \autoref{fig:error_rate} show the individual performance metrics throughout the iterations achieved by three conditions. 
We ran one-way repeated-measures ANOVA on each iteration across all metrics and found that the \method generally led to either comparable or significantly better performances than the other baselines. We denote the significant differences in the figures.

Finally, we analyzed participants' responses to the NASA-TLX questionnaire and found no significant differences across conditions. 
This result is consistent with prior work~\cite{10.1145/3313831.3376244,10.1145/3613904.3642071}, which similarly reported that participants could not easily perceive differences between adaptation methods, even when objective performance differences were present.
The primary reason was that within each adaptation method, different keyboard designs were proposed, making subjective comparison across methods challenging.

\subsection{Findings and Discussion}

\subsubsection{Findings on Performance}
Three methods led to performance improvements across iterations, yet the resulting performances diverged in the early iterations: \method provided an advantage in the second and third iterations.  
TAF's performance is significantly poorer in early iterations, which can be attributed to the limited coverage of its prior models. 
The five priors collected in Phase~1 could not sufficiently capture the diversity of population performance, leading to poor prediction in unexplored regions of the design space. 
In principle, expanding the number of priors could improve coverage and even approach optimal results. 
However, there are two fundamental challenges. 
First, there is no clear principle for predicting how many prior users are sufficient, which makes it difficult to guarantee generalization. 
Second, scaling up the number of priors linearly increases computation during deployment, rendering this approach less practical in real-world scenarios.  
By contrast, \method has the unique strength of not being limited to a finite set of prior users. 
Instead, it can generate unlimited synthetic users by sampling from learned parameter distributions. 
This synthetic population enables extensive offline training before deployment, resulting in more consistent early-iteration performance. 
In addition, the novelty detector allows \method to flexibly adjust the weighting between the NAF policy and the EI values calculated from the current user's model.
This mechanism provides robustness in out-of-distribution cases: when priors are less informative, \method behaves similarly to TAF by relying more heavily on the current user's data, ensuring at least comparable performance.

ConBO showed promising overall performance but was less effective in the earliest iterations, primarily because the first few participants exhibited relatively poor performance during their initial trials. 
The fundamental challenge of ConBO is that it learns gradually across users. 
For the first few participants, the population model contains little information, so its search behavior resembles random exploration. 
Once enough participants are aggregated, however, the model improves steadily. 
This trend is evident when comparing the performance of early and later groups of participants: Group~1 (P1--3) = 0.37 (std = 0.12), Group~2 = 0.61 (std = 0.07), Group~3 = 0.59 (std = 0.19), and Group~4 = 0.63 (std = 0.04). 
These results clearly show that ConBO benefits from accumulating user data, but at the expense of initial participants who experience suboptimal adaptation.  
In contrast, \method does not suffer from this early-stage drawback, as its model is fully trained prior to deployment using synthetic users. 
The group-wise performances further illustrate this stability: Group~1 = 0.54 (std = 0.10), Group~2 = 0.56 (std = 0.12), Group~3 = 0.66 (std = 0.25), and Group~4 = 0.65 (std = 0.03). 
Thus, \method provides better performance for the initial participants compared to ConBO.

To summarize, TAF is limited by finite priors and scalability concerns, ConBO improves gradually but suffers from early-user effectiveness, and \method combines the advantages of prior-based learning with scalable synthetic-user generation to achieve stronger early adaptation. 

\subsubsection{Importance of the novelty detector and EI}

\begin{figure*}[t]
\centering
\includegraphics[width=1\textwidth]{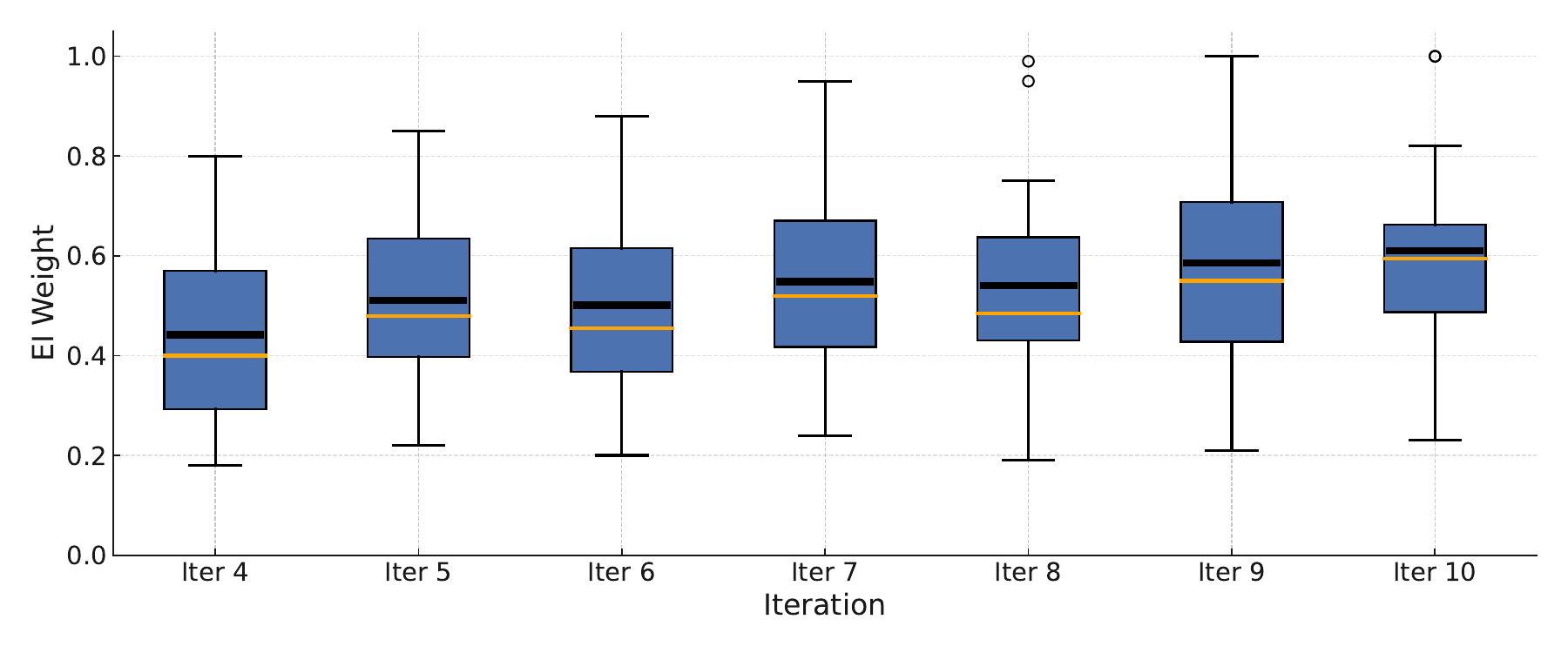}
  \caption{
  The evolution of EI weights ($\lambda_{EI}$) using \method at each iteration. The thicker, black line in each box shows the mean value, and the thin, orange line indicates the median value. Overall, the novelty levels increase across iterations, and participants demonstrate high differences from very novel users (behaviors largely differ from predictions) to well-modeled users (behaviors are highly aligned with predictions). This shows the importance of employing the novelty detector and EI as a fallback acquisition.
  }
 ~\label{fig:ei_weight}
\end{figure*}

We further analyzed the contribution of EI and novelty detector by examining the EI weights of all the participants across iterations 4-10 (at which weighted-sum mechanism performs). 
The EI weights are shown in \autoref{fig:ei_weight}. 
Overall, we observe a gradual upward trend in EI weights, indicating that as the optimization progresses, the optimization increasingly relies on the EI component. 
A closer examination of individual EI-weight trajectories reveals different participant groups that have distinct levels of novelty in performance characteristics: 
A few participants \textbf{well-modeled users}, who consistently show low and stable EI weights (e.g., 0.3–0.5).
These users behave very similarly to the predictive model's expected performance distributions.
A few participants are \textbf{novel users}.
They show consistently high EI weights, often rising rapidly toward the upper range (e.g., 0.7–1.0).
Their performance profiles deviate substantially from what the predictive model expects.
Yet, most participants have \textbf{moderate EI weights} with small oscillations, indicating most participants partly match the model's expectations, which is a natural outcome.

These observations show the importance of the novelty detector and the EI weighting mechanism.
With them, \method can effectively dynamically adjusts how strongly it should rely on learned priors versus current observations, maintaining efficiency without sacrificing personalization across diverse users.

\subsubsection{Qualitative analysis of the user experience with ConBO}

While no significant differences were found in NASA-TLX questions, the first few participants in ConBO shared more frustrated experiences.
As previously analyzed, Group 1 (P1–P3) in ConBO exhibited worse overall performance compared to later groups, indicating that ConBO was still accumulating knowledge and had not yet formed a reliable prior for personalization.
In the interviews, Group 1 participants described the initial keyboards as feeling ``weird,'' with ``impossible ratios'' or being ``way too small for me to type comfortably.'' They also expressed that the system did not seem to adapt meaningfully to their behavior. For example, P2 said, ``I felt the keyboard randomly changes dimensions,'' and P1 noted, ``It did not improve from where I did better.''
In contrast, participants in later groups described a smoother experience. 
Notably, complaints about impossible layouts or misaligned adaptations were no longer observed, nor in the other conditions.
Overall, these observations confirm a clear drawback of ConBO: its performance for early participants is less reliable, as the system is still accumulating experience and has not yet formed a prior for effective optimization.

\section{Discussion}
 
We propose, \conceptnospace,  a novel concept which pre-trains meta-BO using  user models to accelerate adaptation in human-in-the-loop settings.
We further introduce \method, a new optimization method that integrates a deep-learning-based acquisition function, trained via reinforcement learning on synthetic users to support efficient real-time adaptation.
Our synthetic tests and user study validate the efficacy of our approach.
Despite these promising results, several open questions and limitations remain. 

\paragraph{Rethinking personalization: From model-fitting to meta-adaptation}

Traditional model-based personalization requires fitting a parameterized user model to each user, which often involves complex model inverse steps before any adaptation can occur \cite{beaumont2002approximate}. 
Although recent learning-based inference approaches (e.g., amortized inference~\cite{moon2023amortized, moon2024real}) have been proposed to mitigate this challenging, model inference remains a challenging step to support real-time adaptation to the diverse user characteristics.
In contrast, \concept adopts a different philosophy: we do not seek to model any single user perfectly.
Instead, we train the optimizer on a broad distribution of synthetic users, enabling it to learn adaptation strategies that generalize across populations. 
At deployment, no explicit user modeling or inference is needed; the system simply adapts based on observed performance, guided by the meta-learned policy. 
This shift from high-fidelity modeling toward population-level diversity and meta-adaptation offers a more scalable and robust path for real-time personalization.

\paragraph{Re-purposing user models} 

User models in HCI have traditionally been developed as descriptive tools mainly for characterizing and predicting human performance, but not meant for interface generation and adaptation.
Thus, even for well-studied behaviors such as pointing or typing, available models were rarely fitted comprehensively across devices or modalities, because there was little downstream use for such parameter values. 
This gap creates a potential challenge for HOMI: our workflow benefits most when reliable models and parameter ranges already exist, enabling us to bypass the parameter-fitting step.
Our \concept and \method introduce a fundamentally different purpose for user models: transforming them from passive analytic tools into active enablers of interface optimization, adaptation, and even generation.
This re-positioning creates an actionable incentive for the community: rather than treating user models as isolated theoretical artifacts, we encourage documenting, validating, and sharing their parameter values across interactions, devices, and modalities.
A collaboratively maintained gallery of models and parameter distributions. 
That is, not just equations, but reusable parameter sets, would unlock scalable synthetic-user–driven optimization and significantly broaden the applicability of HOMI-style workflows across HCI.

\paragraph{Extending to other applications}
Our approach is built upon the broad framework of BO and RL, which have shown strong scalability and generalizability across applications. 
In future work, we will apply the \concept framework to other interactive systems where well-established performance or cognitive models exist.
Promising examples include menu optimization, gaze-based interfaces, gesture input, and general scenarios where task-specific models can be leveraged to generate meaningful synthetic users and enable pretraining.
We can then further validate the flexibility of our approach and demonstrate its potential as a general solution for scalable, adaptive interface optimization.

\paragraph{Utilizing advanced models to generate synthetic users}
\concept is grounded in the use of parameterized user models to simulate synthetic users and enable large-scale pretraining.
While this provides a scalable and principled way to approximate diverse user behaviors, it requires established models for a given interaction task.
Such models might not be available for more complex interactions that involve rich cognitive processes, sequential decision-making, or dexterous motions.
Here, we see an opportunity for the use of more advanced modeling approaches that have recently emerged in HCI. Specifically, advances in Computational Rationality and reinforcement learning–based modeling of cognitive and motor processes ~\citep{oulasvirta2022, simulationHCI, 10.1145/3708359.3712114} offer promising alternatives for simulating realistic human behavior in such scenarios.
When combined with biomechanical models that provide realistic musculoskeletal dynamics \cite{ikkala2022breathing, moon2024real, moon2024real, 10.1145/3491102.3501992, ikkala2022breathing, 10.1145/3654777.3676452}, this approach offers a powerful path toward general-purpose human simulators. 
Unlike traditional analytical models that are tied to specific tasks, these reinforcement learning-based models can learn to reproduce a range of sensorimotor behaviors through learning. 
This opens the door to synthetic users that can generalize across tasks, contexts, and devices, further broadening the applicability of \concept.
Additionally, generative models, such as variational autoencoders, diffusion models, or large language models, are increasingly used to model human-like behavior from data, and could be adapted to synthesize user interaction patterns in tasks where explicit models are lacking \cite{li2024personalized, park2025llm}.
Future work could incorporate these approaches to expand the scope of \conceptnospace, or explore hybrid user simulation strategies that blend analytical models with data-driven user representations.
These directions can significantly broaden the applicability of our framework across a wider spectrum of HCI domains.

\paragraph{Generalization beyond synthetic users}

Although our novelty detector provides a fallback mechanism for handling out-of-distribution users, the overall performance of \method still depends on how well the synthetic user distribution approximates real-world variability.
Moreover, the sampling of synthetic users presents an inherent trade-off.
Sampling a broad and diverse range of synthetic users allows \method to learn generalizable strategies across many user profiles, but this may come at the cost of reduced performance for any specific user.
In contrast, sampling from a narrow range enables the optimizer to acquire more specialized strategies tailored to users that are more aligned with the synthetic ones, but at the risk of having a higher chance to encounter users outside of the training distribution.
Balancing this exploration–exploitation trade-off in user simulation is a key open question.
Future directions may include adaptive sampling, curriculum-based simulation, or progressive refinement of the synthetic user population based on deployment feedback.
Such approaches could further enhance the generalization capacity of \method in real-world human-in-the-loop scenarios.

\paragraph{Development of other \concept methods}
While this paper presents \method as one realization of the broader \concept framework, the framework itself is general and can support many alternative instantiations.
One especially promising direction is to incorporate online user data into the meta-learning loop; that is, incorporating continual HILO \cite{liao2025continual} into our workflow.
Currently, the meta-optimizer is trained exclusively on synthetic users generated from established models.
However, as real user interaction data accumulates over time, integrating this data into the training process could allow the optimizer to continuously improve and better adapt to emerging usage patterns and behaviors.
Such lifelong learning mechanisms would allow \conceptnospace-based systems to evolve beyond their initial modeling assumptions and progressively refine their understanding of user diversity.
Future \concept methods could also explore alternative optimization strategies, such as moving beyond weighted-sum aggregation in multi-objective optimization and explicitly model Pareto frontiers, allowing users or systems to select among optimal trade-offs post hoc~\cite{paretoAdapt}.
Further, future research could investigate alternative acquisition architectures, such as transformer-based models or hybrid schemes that combine learned and analytical acquisition components.
These extensions would further enhance the flexibility and impact of \concept across a broader range of real-world HCI scenarios.

\section{Conclusion}

We introduce \conceptnospace, a novel framework for HILO that leverages model-informed pre-training to accelerate real-time interface adaptation. 
By synthesizing large populations of simulated users from user models, \concept enables the training of powerful optimizers prior to deployment, bridging the gap between traditional model-based design and purely interactive approaches.
To demonstrate the framework, we proposed \methodnospace, a concrete method that integrates a deep reinforcement learning–based acquisition function, dynamic multi-objective adaptation, and a novelty-aware fallback mechanism.
Together, these components enable efficient and robust optimization across diverse users and changing objectives.
Our results suggest that this new concept and approach jointly offer a scalable and generalizable path for future personalization systems: one that moves beyond per-user modeling and instead learns population-level adaptation strategies that generalize in deployment.
We believe this work opens up a new direction for optimization in HCI, one that treats user models not just as analytical tools, but as generative engines for training adaptive, intelligent design systems.
Finally, we hope this work provides a new lens for bridging \emph{in vitro} (in simulation) with \emph{in situ} (with real users) optimization, enabling more streamlined and adaptive future interfaces.

\section{Open Science}

\noindent
\method is released at \href{https://github.com/yichiliao/homi}{https://github.com/yichiliao/homi}. We hope that our implementation can encourage other researchers to build other applications based on our proposed workflow and approach.

\begin{acks}
The authors sincerely thank Aakar Gupta, Ruta Desai, Antti Oulasvirta, and Aleksi Ikkala for their insightful discussions. Yi-Chi Liao was partially supported by the ETH Zurich Postdoctoral Fellowship Programme. Hee-Seung Moon is supported by the National Research Foundation of Korea (RS-2025-00521470).
\end{acks}

\bibliographystyle{ACM-Reference-Format}
\bibliography{sample-base}
\appendix
\clearpage
\newpage

\section{Appendix: Overview of Simulated Tests}

In this appendix, we describe the synthetic experiments conducted to validate the effectiveness of \methodnospace. These experiments are designed to address the following three goals:

\begin{itemize}
    \item \textbf{Goal 1: Compare performance against baselines.} We benchmark \method against two established HCI optimization approaches: standard Bayesian Optimization (BO) and Transfer Acquisition Function (TAF).
    \item \textbf{Goal 2: Validate dynamic objective weighting.} A key feature of \method is its ability to adapt to varying weights across multiple design objectives during deployment. We test this capability under shifting weight settings.
    \item \textbf{Goal 3: Evaluate robustness to novel users.} Another contribution of \method is its mechanism for handling out-of-distribution users through a novelty-aware fallback strategy. We examine its efficacy in these scenarios.
\end{itemize}

To evaluate these goals, we conduct two synthetic tests. The first is a standard black-box optimization problem (Sphere function), commonly used for evaluating BO methods. The second is a semi-realistic simulation designed to mirror our target use case: optimizing mid-air keyboards for users with diverse performance profiles. In this simulation, we generate synthetic users with varying typing behaviors and aim to optimize keyboard layouts per individual.

\subsection{Optimization Approaches (Conditions)}

We define a set of optimization approaches, or ``conditions,'' which are held consistent across both synthetic tests.

\paragraph{1. \methodnospace (full).}
This is the complete version of our proposed method, featuring both: (1) dynamic handling of multi-objective weights through explicit input, and (2) fallback support for novel users via the combination of EI and a novelty detector.

\paragraph{2. \method without explicit multi-objective input.}
This variant of \method is trained to handle different objective weightings but does not receive explicit weight vectors as input during training or deployment. This tests the importance of conditioning the policy directly on objective weights.

\paragraph{3. \method without novel user fallback.}
This version excludes the components responsible for handling novel users (i.e., no EI fallback and novelty detection). It relies solely on the neural acquisition function (\texttt{naf}) trained on synthetic users.

\paragraph{4. Standard BO}
This is the typical Bayesian Optimization setup, using Gaussian Processes and Expected Improvement as the acquisition function. It does not utilize any pretraining or meta-learning. This condition serves as a standard baseline in HCI for human-in-the-loop optimization.

\paragraph{5. Transfer Acquisition Function (TAF)}
TAF is a meta-learning approach widely used in prior HCI work~\cite{wistuba2018scalable, 10.1145/3613904.3642071}. It builds a library of prior user models, each is as a standalone Gaussian Process (GP) trained on that user's optimization data. During deployment, all prior models independently generate acquisition values for the candidate designs, and these values are aggregated—typically via a weighted sum—to guide the optimization for a new user.
A key distinction between TAF and all variants of \method lies in scalability. While TAF relies on a fixed set of prior users and its computational cost increases linearly with the size of the prior model library, \method is trained on an arbitrarily large population of synthetic users but retains constant complexity. This enables \method to scale more efficiently with increasing amounts of data.

Note that we did not include the vanilla NAF as a separate condition because it essentially operates as our \method without both the explicit multi-objective input and the novel user fallback mechanism. 
The differences between \method and NAF are already captured in the comparisons between Condition 1 vs Conditions 2 and 3.

\subsection{Generating a Group of Synthetic Users}

We follow standard procedures for evaluating meta-learning and transfer optimization methods~\cite{volpp2019meta, wistuba2018scalable, 10.1145/3613904.3642071}, simulating a population of synthetic users who share core behavioral structure but exhibit individual variability.
Each synthetic user is represented as a unique function, derived by modifying a shared base function through controlled transformations. We refer to these personalized variants as user functions. These are divided into a training group (used to train the meta-optimizer) and a test group (used to evaluate generalization at deployment).

For the first test, we determine a base function using a common test function for optimization tasks. 
Then, each user function is created by shifting the function (translating the input values before passing them to the function). 
The magnitude of the shifts is uniformly sampled from a distribution within a specified range, adding diversity among the users. Mathematically, the shift is represented as $x_n' = x_n + \delta_n$, where $x_n$ is the original input, $x_n'$ is the shifted input, $n \in {[1, N]}$ represents the parameters, and the shift amount $\delta_n \sim U(-\frac{shift\_range}{2}, \frac{shift\_range}{2})$. This step simulates different user responses to the same design.
We also scale the output of the function by a scalar factor, which is also sampled from a uniform distribution. This factor, denoted as $\mathcal{S} \sim U(1-\frac{scale\_range}{2}, 1+\frac{scale\_range}{2})$, introduces further diversity in users' performance level.
Through the combinations of these randomly sampled shift and scale, we generate a diverse set of user functions that preserve the structure of the base function while simulating realistic inter-user differences.

For the second test, we employ a user model based on existing literature, which has 6 user parameters. 
We sampled the parameter values from the range provided by the existing literature; a unique parameter assignment represents a specific user with a specific user performance given a design.
More details will be provided in the second synthetic test (\autoref{sec:test2}).

\subsection{General Procedures in the Synthetic Tests}
Our synthetic tests follow a consistent structure consisting of three stages: training, testing, and evaluation on novel users.

\paragraph{Step 1: Training}

We first train all \method variations and TAF using a collection of prior user functions. For TAF, the training process involves fully optimizing each prior user function to build a set of GP models—one per user. To enable TAF to handle dynamic objective weights, we follow a prior approach~\cite{10.1145/3613904.3642071} where multi-objective optimization is performed during training. This results in GP models that can predict multiple objective values, allowing the optimizer to accommodate dynamic weight assignments during deployment. We train TAF with 15 prior user functions, which is consistent with earlier studies. Increasing the number of users further leads to prohibitively long computation times at deployment, as TAF's runtime scales linearly with the number of stored models.

For all \method variants, we dynamically generate synthetic users and continue training until a fixed number of training steps is reached ($80000$ steps for both synthetic tests). The three \method variants share the same training procedure, with the only difference being that Condition 2 (\method without explicit multi-objective input) is trained on multiple objective weight configurations but does not receive the weight information as part of the model input. Additionally, all collected $(x, y)$ pairs during training are used to train the Bayesian Neural Network–based novelty detector. Notably, the novelty detector is trained with the objective weights as part of the input, enabling it to condition its uncertainty estimation accordingly. However, Condition 3 (\method without novelty detection) does not include this module and skips this training step.

\paragraph{Step 2: Testing}

Next, we sample 20 test functions, each representing a new synthetic user with distinct characteristics. 
For each test function, the objective weights are randomly sampled from a uniform distribution. These functions are also shifted and scaled as described earlier to ensure diversity. 
Each condition is then deployed independently on these test functions to evaluate its performance with a budget of 20 iterations.

\paragraph{Step 3: Testing on Novel Users}

To evaluate the robustness of each condition under distributional shift, we construct a separate set of “novel” user functions. These are generated using parameter values that lie completely outside the range observed during training. All optimization conditions are then tested on these novel users to assess their ability to generalize to unseen behaviors.

\section{Synthetic Test 1: Double-Sphere Functions}

In this synthetic test, we define a composite 2D objective function to evaluate the optimizer's ability to handle dynamic objective trade-offs as well as coping with novel users. 
Specifically, two sub-tests are conducted to evaluate two aspects of \methodnospace. 
The first sub-test (denoted as test 1.1) evaluates the importance of weight information, and the second sub-test (test 1.2) validates the importance of incorporating the typical expected improvement acquisition funciton. 

\subsection{Function Details}

The function has two input parameters and two output objectives, each modeled as a 2D Sphere function centered at different locations. 
Specifically, the first Sphere function is centered at $(0.4, 0.4)$, while the second is centered at $(0.6, 0.6)$. 
Each Sphere function is defined as $y = 1 - \sum_{i=1}^2 (\dot{x}_i - X_i)^2 \times \gamma$, where $X_i$ is the center coordinate and $\gamma = 8$ controls the sharpness of the peak. 

The final objective value is computed as the weighted sum of these two Sphere functions, with weights dynamically assigned during deployment. 
This creates a setting where the global optimum shifts depending on the weight configuration: when the first objective is prioritized, the optimal region lies near $(0.4, 0.4)$; when the second is prioritized, it shifts toward $(0.6, 0.6)$. 
To simulate user variability, we directly apply random shifts to the input space $\mathbf{x}$ in both dimensions, as well as scaling to the final weighted-sum output of the two sub-functions. 
Specifically, each synthetic user is defined by a unique shift vector and scaling factor, applied consistently across both objective components. 
This formulation introduces variation in both user preference (by altering the location of optima) and sensitivity (by modulating the steepness of the reward surface), while preserving the underlying structure of the task. 


\subsection{Configurations of each Condition}
\label{sec:test_configuration}

Here, we detailed the configurations of different conditions. 

\paragraph{1. \method}
We set the \texttt{naf} as a six-layer fully connected neural network, with each hidden layer containing 256 nodes and Tanh as the activation function. The \texttt{naf} is trained using Proximal Policy Optimization (PPO) \cite{schulman2017proximal}, where the critic network shares the same architecture as the actor (i.e., \texttt{naf}). 
The actor and critic learning rates are set to 0.0003 and 0.001, respectively. PPO is run for $K = 10$ epochs with a clipping parameter of 0.15.
To encourage generalization across varying objective weightings, the objective weights are randomly sampled during training. To be more specific, the weight on the first Sphere function, $\alpha$, is drawn uniformly from $[0, 1]$, and the weight on the second is set to $1 - \alpha$. This two-dimensional weight vector $(\alpha, 1 - \alpha)$ is also included as part of the input to the \texttt{naf}, allowing the acquisition function to adapt its strategy based on the current configuration.
Note, when deploying on the test functions, the weights are given to \method as a part of the input.
Finally, the pre-defined novelty threshold $\tau$ is set to be 0.1. 
We denote this method as \method in the rest of this test.

\paragraph{2. \method without weights information (NAF-w/o-Weight)}
This condition is very similar to the full version of \methodnospace. 
The model setting, hyperparameter used in training, and training procedure are identical to \methodnospace. 
The only difference is that this method is not informed by the information of the objective weights. 

\paragraph{3. \method without EI (NAF-w/o-EI)}
This condition is very similar to the full version of \methodnospace. 
The only difference is that there is no additional GPs constructed for the new function, and thus, there is no EI or novelty detector. The process is fully dominated by NAF all the time.

\paragraph{4. Transfer Acquisition Function (TAF)}
This condition is Transfer Acquisition Function (denoted as TAF). 
This approach is a weighted-sum method; a series of GP models will be created and stored prior to deployment.
We follow the previous works (\cite{10.1145/3613904.3642071}) to configure TAF; 15 prior GPs are gathered, each is trained from a randomly sampled double-Sphere function, with the objective weights also randomly sampled. 
These GPs are constructed via standard BO, with 10 initial random sampling with 15 optimization steps. 
Similar to \citet{10.1145/3613904.3642071}, we configure TAF with decaying model weights on the previous GPs, which allows the new GP model to gradually gain control over the optimization process.
Specifically, $\alpha_1$ is set as 4 and $\alpha_2$ is set as 0.2. 

\paragraph{5. Standard Bayesian optimization (BO)}
The final condition is standard Bayesian optimization (BO). 
We used Expected Improvement as the acquisition function and a typical Single-Task GP implemented in the BoTorch library as the surrogate model\footnote{\url{https://botorch.org/}}.
We set BO with 6 random explorations followed by 14 optimization steps.

\subsection{Result of Test 1.1 (Evaluating Weight Information)}

\begin{figure*}[h!]
\centering
\includegraphics[width=1\textwidth]{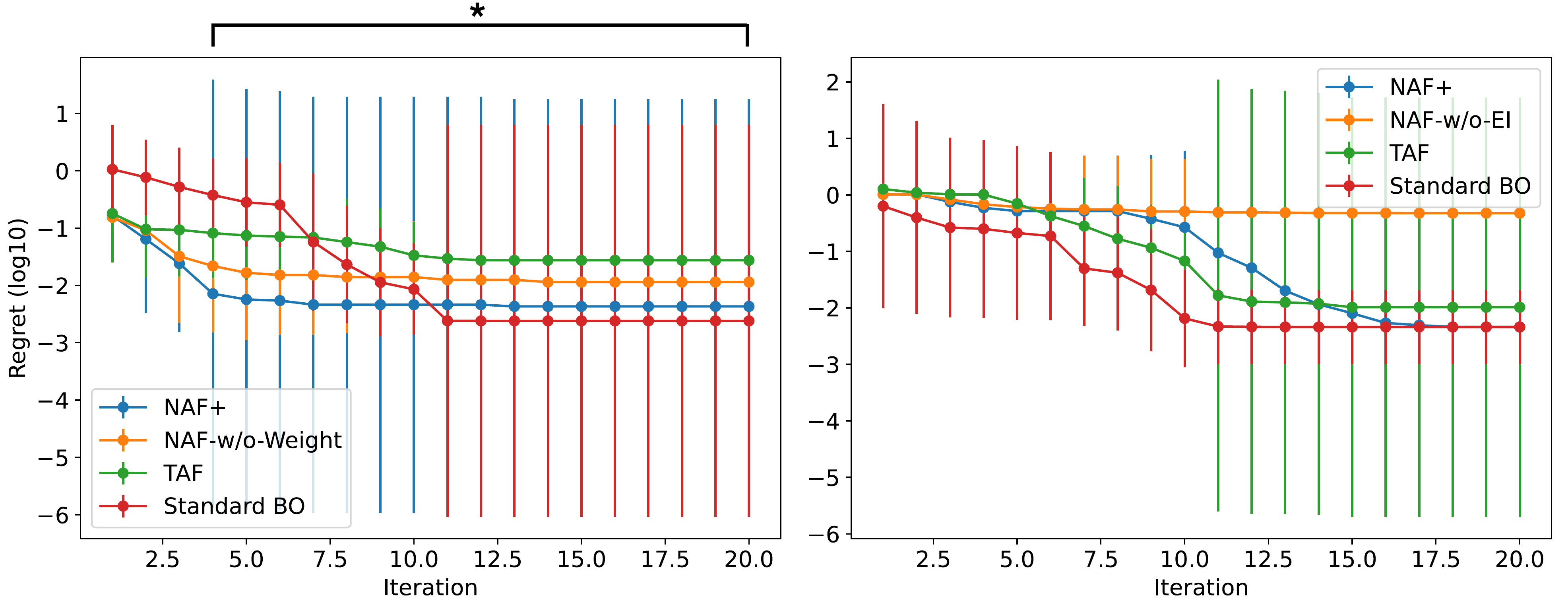}
  \caption{
  \rev{The results of Test 1 (double-sphere functions): Test 1.1 (evaluating weight information) is shown in left, and Test 1.2 (evaluating expected improvement) is shown in right.} The value plot here is the $log_{10}$ of the regret. In Test 1.1, the * signs denote the iterations (5-20) where a significant difference between NAF+ and NAF-w/o-Weight was found. The results show that \method benefits from the weight information and thus can achieve better performance than NAF-w/o-Weight by a more informed optimization strategy. Meanwhile, the novelty detector allows \method to rely on its GP and EI to address tasks that are outside of the training distribution, whereas NAF-w/o-EI can not adapt to novel tasks.
  }
 ~\label{fig:test1}
\end{figure*}

Test 1.1 is designed to specifically evaluate the value of weight information.
To this end, we sampled 20 test functions from the exact same range as the one used for training. 
Since the test functions are fully overlap with the training range, we exclude the condition with novel detection and additional EI (NAF-w/o-EI) from this test 
The log regret results are presented in \autoref{fig:test1} left panel. 
Overall, we can observe that \method started from an overall better performance and constantly outperformed other meta-BO approaches, such as \textit{NAF-w/o-Weight} and TAF.
Also, while TAF showed promising performance (most regret is less than $10^{-1}$), it still performed worse than \method and NAF-w/o-Weight. 
The primary reason is that \method is trained by a nearly unlimited number of synthetic agents, allowing it to fully generalize across new tasks. 
On the other hand, TAF is based on a limited set of prior synthetic users. 
In addition, within the first 10 iterations, all meta-BO approaches demonstrated better performance than standard BO, which starts from scratch. This highlights the benefit of prior experience.
To further analyze the benefit of incorporating the weight information, we run independent-samples t-tests to specifically compare the resulting performance of \method and NAF-w/o-Weight at each iteration. 
We found significant differences in iteration 5-20, all $p<0.05$, confirming that NAF overall outperforms NAF-w/o-Weight.

\subsection{Result of Test 1.2 (Evaluating Expected Improvement)}

Test 1.2 evaluates the benefits of incorporating EI and a novelty detector to handle out-of-distribution cases.
This time, we sampled 20 test functions outside the training range, which is a possible case in a real-world scenario where real users may not be fully aligned with our models.
To that end, we directly shifted the center of the two Sphere functions from $(0.4, 0.4)$ and $(0.6, 0.6)$ to $(0.3, 0.3)$ and $(0.7, 0.7)$. 
This ensures the sampled functions will be different from the training functions, but not too distant. 
We include the condition with novel detection and additional EI (NAF-w/o-EI) from this test, but exclude the condition of weight information (NAF-w/o-Weight) since it is not a part of the main comparison.
The log regret results are presented in \autoref{fig:test1}, right. 
Without incorporating EI based on the newly constructed GP, the performance of NAF did not make significant improvements throughout the iterations. 
This is expected, as the test functions are out of the training distribution.
On the other hand, \method did not start with a good performance due to the out-of-distribution test case.
However, its performance gradually catches up with the other baselines, demonstrating the necessity of incorporating EI in real-world scenarios. 

\subsection{Summary of test 1}

Test 1.1 answered \textbf{goal 1 (comparing performance against baselines)}: we observe \method and NAF-w/o-Weight generally outperformed TAF owing to its rich training data. 
\method and NAF-w/o-Weight also significantly outperformed Standard BO in the first few iterations. 
Test 1.1 also answered \textbf{goal 2 (validating dynamic objective weight)}: we found that overall, \method converges faster than NAF-w/o-Weight, owing to its capability to take objective weight as input and adjust its search policy. 
Test 1.2 responded to \textbf{goal 3 (evaluating robustness to novel users)}: we showed that when facing out-of-distribution users, the novelty detection with additional EI ensures robust convergence. Meanwhile, NAF-w/o-EI can not effectively address such scenarios.

\section{Synthetic test 2: Simulating softkeyboard typing}
\label{sec:test2}
In the second synthetic test, we develop a typing user model that closely mirrors the target behavior of our real user study. 
This model predicts user performance on a given keyboard design, enabling us to evaluate design quality by simulating the act of typing a sentence on the sampled keyboard. 
As in the previous test, we conduct two sub-tests: one to assess the importance of including weight information as part of the model input, and another to evaluate the benefit of incorporating Expected Improvement (EI).

\subsection{Model Details and Generating Synthetic Users}

In this test, a keyboard design is defined by its spatial dimensions, including height and width. 
The user model predicts two key performance metrics: (1) \textit{typing error}, defined as the probability of pressing an incorrect key, and (2) \textit{movement time} between consecutive key presses. 
The model is composed of two analytical components: Fitts' Law~\cite{mackenzie1992extending} for modeling movement time, and the Dual Gaussian Distribution (DGD) model~\cite{10.1145/2984511.2984546} for predicting touch error on touchscreen keyboards. 
Further implementation details are provided in Section 5 of the main paper.

To generate synthetic users, we assign values to the model parameters $[a, b, \alpha_x, \sigma_{\alpha_x}, \alpha_y, \sigma_{\alpha_y}]$ by sampling from distributions informed by prior work. 
\citet{10.1145/2470654.2466180} report Fitts' Law parameters fitted to touchscreen typing as $a = 144.3$ and $b = 75.636$, with the unit being milliseconds. 
We use these values as the means of two independent Gaussian distributions (each with a standard deviation of 15) and sample from them to generate user-specific values of $a$ and $b$.
Here, we emphasize that this is a synthetic task; the exact parameter values vary for different input modalities and keyboard settings, and here we only take the available model values to demonstrate the efficacy of our approach.

For the touch error model, we adopt the generic parameterization from the DGD model~\cite{10.1145/2984511.2984546}, setting the means of the Gaussian distributions to $\alpha_x = 0.0075$, $\sigma_{\alpha_x} = 1.296$, $\alpha_y = 0.0108$, and $\sigma_{\alpha_y} = 1.153$. 
The corresponding standard deviations are arbitrarily set to $0.001$, $0.01$, $0.001$, and $0.01$, respectively. 
Each parameter is sampled independently, resulting in a diverse population of synthetic users with varying motor behavior and touch precision profiles.

\subsection{Configurations of each Condition}

All the conditions remain the same as in the previous synthetic test (\autoref{sec:test_configuration}).

\subsection{Result of Test 2.1 (Evaluating Weight Information)}

Similar to the previous test, Test 2.1 is designed to evaluate the value of the weight information. 
We sampled 20 test functions from the exact same range as the one used for training. 
We again exclude the condition with novel detection (NAF-w/o-Weight) from this test as there will not be novel functions. 
The log regret results are presented in \autoref{fig:test2}, left panel. 
\method and NAF-w/o-Weight start from an overall better performance than the others. 
Especially, \method converges to optimal performance around the second iteration, and NAF-w/o-Weight converges around the fifth iteration, showing \method still benefits from the additional weight information as a part of the input. 
We observed that TAF with 15 prior models struggled to improve significantly in this case.
This might be because of the higher number of parameters (2 in Test 1 and 6 in Test 2). 
To cover this number of dimensions, potentially more models will be required. 
Furthermore, same as the previous test, Standard BO starts from a worse performance and gradually converges toward optimal performance. 
Finally, to further analyze the benefit of incorporating the weight information, we run independent-samples t-tests to specifically compare the resulting performance of \method and NAF-w/o-Weight at all iterations. 
We found significant differences in iterations 1-4, all $p<0.05$. 
This shows that the weight information supports more efficient adaptation in the earlier iterations.

\subsection{Result of Test 2.2 (Evaluating Expected Improvement)}

\begin{figure*}[h!]
\centering
\includegraphics[width=1\textwidth]{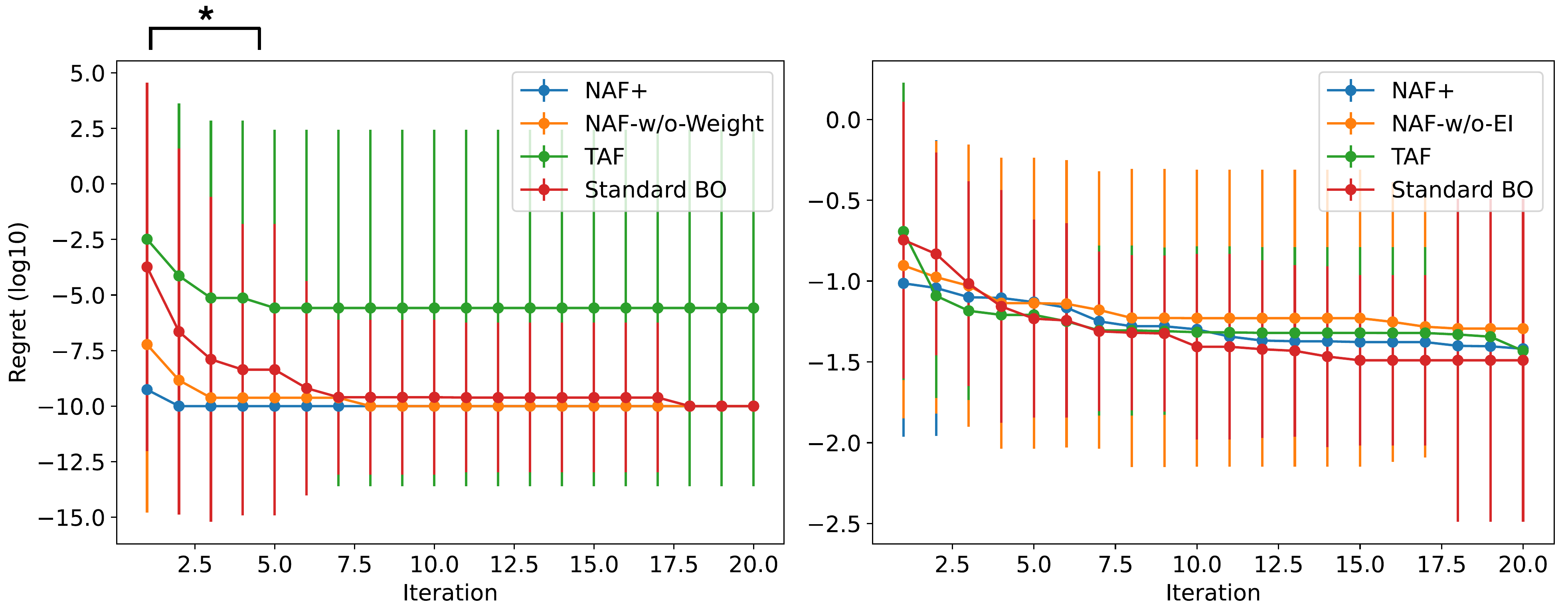}
  \caption{
  \rev{The results of Test 2 (simulating softkeyboard typing): Test 2.1 (evaluating weight information) is shown in left, and Test 2.2 (evaluating expected improvement) is shown in right.} The value plot here is the $log_{10}$ of the regret. In Test 2.1, the * signs denote the iterations (1-4) where a significant difference between NAF+ and NAF-w/o-Weight was found. The results show that \method benefits from the weight information and thus can achieve better performance than NAF-w/o-Weight in earlier iterations. On the other hand, \method can achieve better results compared to NAF-w/o-EI when facing novel tasks, particularly in later iterations.
  }
 ~\label{fig:test2}
\end{figure*}

Test 2.2 evaluates the value of merging EI and the novelty detector. 
To create a scenario where the new user is completely outside of the training range, we significantly shift the range of sampling the model parameters. 
The mean for sampling synthetic users are set as $[180, 120, 0.0145, 1.6, 0.03, 2.0]$ for $[a, b, \alpha_x, \sigma_{\alpha_x}, \alpha_y, \sigma_{\alpha_y}]$, respectively, and the standard deviation remains the same.
These ranges are significantly different from the parameter settings for training, ensuring all sampled users are novel. 
We sampled 20 test functions from the test range. 
The log regret results are presented in \autoref{fig:test2}, right panel. 
We find that \method and NAF-w/o-EI start from overall better performance, but NAF-w/o-EI does not improve significantly throughout the iterations, while \method continually improves. 
This, again, highlights the potential issue of not incorporating an additional EI to handle out-of-distribution cases:  \texttt{naf} can not generalize to other cases once it is fully trained to a certain range. 
TAF and standard BO perform well in this task. 
However, they start from a worse performance. 
BO, especially, requires 4 to 5 iterations to catch up with other conditions.

\subsection{Summary of Test 2}

Test 2.1 addresses \textbf{goal 1 (comparing performance against baselines)}. We observe that \method and NAF-w/o-Weight consistently outperform TAF, likely due to the higher dimensionality of the input space in this test. TAF, relying on a limited set of prior models, would require a larger number of samples to adequately cover the expanded design space. 
\method and NAF-w/o-Weight also significantly outperform standard BO in the early iterations, demonstrating the benefits of leveraging prior knowledge.
Test 2.1 also supports \textbf{goal 2 (validating dynamic objective weighting)}. Compared to NAF-w/o-Weight, \method achieves faster convergence by conditioning its acquisition strategy on the provided objective weights, resulting in higher sample efficiency.
Test 2.2 addresses \textbf{goal 3 (evaluating robustness to novel users)}. We find that NAF-w/o-EI struggles to adapt when users fall outside the training distribution, while \method—equipped with a novelty-aware EI fallback—successfully recovers and improves performance over time.

In summary, both Test 1 (a controlled synthetic benchmark) and Test 2 (a more realistic simulation grounded in HCI literature) converge on the same conclusion: \method, through its use of explicit weight input and novelty-aware EI, achieves greater sample efficiency and robustness across a wide range of user profiles. 
The positive outcomes of Test 2 further motivate our user study, which targets typing performance on mid-air keyboards, highly relevant to Test 2 -- a touch-based keyboard for mobile devices.

\end{document}